\documentclass[11pt]{article}
\usepackage[preprint]{acl}

\usepackage{times}
\usepackage{latexsym}

\usepackage[T1]{fontenc}
\usepackage[utf8]{inputenc}

\usepackage{microtype}

\usepackage{inconsolata}

\usepackage{graphicx}

\usepackage{booktabs}
\usepackage{amsmath}
\usepackage{amssymb}
\usepackage{multirow}

\usepackage{cleveref}

\title{The Effect of Document Selection on Query-focused Text Analysis}

\author{Sandesh S Rangreji \\
  Johns Hopkins University \\
  \texttt{srangre1@jh.edu} \\\And
  Mian Zhong \\
  Johns Hopkins University \\
  \texttt{mzhong8@jh.edu} \\\And
  Anjalie Field \\
  Johns Hopkins University \\
  \texttt{anjalief@jhu.edu} \\}

\begin{document}
\maketitle

\begin{abstract}
Analyses of document collections often require selecting what data to analyze, as not all documents are relevant to a particular research question and computational constraints preclude analyzing all documents, yet little work has examined effects of selection strategy choices. We systematically evaluate seven selection methods (from random selection to hybrid retrieval) on outputs from four text analyses methods (LDA, BERTopic, TopicGPT, HiCode) over two datasets with 26 open-ended queries.
Our evaluation reveals practice guidance: semantic or hybrid retrieval offer strong go-to approaches that avoid the pitfalls of weaker selection strategies and the unnecessary compute overhead of more complicated ones.
Overall, our evaluation framework 
establishes data selection as a methodological decision, rather than a practical necessity, inviting the development of new strategies.
\end{abstract}

\section{Introduction}
\label{sec:introduction}

Open-ended analysis of large text corpora
fundamentally requires down-selecting data for analyses to be feasible.
When text is analyzed manually, data must be sufficiently small to enable iterative rounds of coding (e.g., \citealt{Yakubi_Gac_Apollonio_2022,fitzpatrick_2022_corporate,klein2024inside}).
Automated methods like Latent Dirichlet Allocation (LDA) \citep{blei2003latent}, BERTopic \citep{grootendorst2022bertopic}, and related approaches \cite{roberts2014structural,srivastava2017autoencoding} allow for analyzing larger datasets than manual annotation (1000s-100Ks of documents), but still require substantial time and compute to scale.

The recent development of LLM-powered methods for corpus analysis, like TopicGPT \citep{pham-etal-2024-topicgpt}, LLooM \citep{lam2024lloom}, and HiCode \citep{zhong-etal-2025-hicode} inherit and exacerbate the same data selection challenges.
While these methods show evidence of supporting more nuanced analyses than their predecessors, compute needs or API costs make exhaustive analysis prohibitively expensive and unnecessarily wasteful: most documents in large corpora are irrelevant to any specific research question.

Current proposed pipelines, however, fail to investigate the impacts of data selection. \citet{pham-etal-2024-topicgpt} and \citet{lam2024lloom} rely on uniform random samples of 200-1000 documents, despite acknowledging that small data fractions may not be representative and ``concepts could differ if run on a significantly larger dataset'' \citep{lam2024lloom}.
\citet{zhong-etal-2025-hicode} use the same keyword queries as qualitative researchers for data filtering, assuming well-chosen queries capture relevant documents.
In general, the document selection process is treated as a practical necessity rather than a methodological decision that fundamentally shapes which themes can be discovered.

In this work, we address this gap by investigating how data selection strategies impact
final outputs of text analysis pipelines.
We focus on settings where the user or researcher has a specific research question or query of interest, for which not all documents in the collection are relevant.
We then investigate how data selection with various degrees of relevance control, from no relevance bias (random selection) to strong relevance control (semantic retrieval with query expansion),
influence the final topics or themes discovered.

We specifically compare seven data selection strategies, including commonly-used information retrieval~(IR) mechanisms (keyword search, semantic search, hybrid search, and more advanced modifications).
While we leverage retrieval methods as a data selection strategy, the analytical task differs fundamentally from traditional information retrieval. IR evaluation focuses on ranking precision for small retrieval sets, typically top-1 to top-20 documents, to satisfy specific information needs. In contrast, corpus analysis requires selecting hundreds or thousands of documents from million-scale corpora.
Thus we run selected data through four different analysis methods (LDA, BERTopic, TopicGPT, and HiCode) and evaluate the identified topics for their relevance (query-alignment), diversity, and relative coverage. We report results over two public health datasets: TREC-COVID (171K biomedical articles) and Doctor-Reviews (171K physician reviews), with 15 and 11 expert-designed queries respectively.

Our primary contributions include (1) a framework for investigating the impact of data selection choices in corpus  analyses (2) systematic evaluation of data selection strategies, and (3) recommendations for which data selection strategy to use, depending on analysis goals and compute resources.
We find that hybrid retrieval methods that combine lexical and semantic search are a strong go-to approach in most settings, offering a good balance of relevance and diversity in final themes, while advanced modifications like query expansion provide little added benefit, and random sampling or keyword-based metrics are generally dispreferred. 
This work moves document selection for corpus analysis from an unexamined practical necessity to a deliberate methodological choice with empirically validated implications for discovery quality.

\section{Related Work}
\label{sec:related}
\paragraph{Manual Data selection}
Manual analyses of text archives typically select what data to analyze through snowball search methods, including iterative keyword searches. Examples includes studies of the Truth Tobacco Industry Documents Library \citep{anderson2011tobacco} and the Opioid Industry Documents Archive UCSF-JHU \citep{caleb2022opioid,klein2024inside}, which conduct searches using pre-implemented black-box search engines with no description of their underlying algorithms.
While iterative searches can be quite thorough, involving manually screening hundreds of documents per iteration and refining queries through multiple rounds, achieving ``saturation'' (receiving identical documents from varied query formulations) often proves infeasible within research timelines \citep{anderson2011tobacco}, thus motivating a more principled approach to understanding the trade-offs of search choices.

\paragraph{Data selection for topic modeling}
When the selected data will be analyzed using an automated approach \citep{blei2003latent,grootendorst2022bertopic,pham-etal-2024-topicgpt,zhong-etal-2025-hicode} rather than manually, the data selection step is also typically automated, with randomly sampling to stay within compute and budget constraints being the default approach \citep{lam2024lloom,pham-etal-2024-topicgpt}.
\citet{maier2020document} provide the most systematic examination of sampling effects on topic modeling, demonstrating that random sample-based topic models closely resemble full-corpus models when samples exceed approximately 10,000 documents
% measuring coherence, reproducibility, and model similarity across news articles, websites, and tweets.
However, their work exclusively examines random sampling at varying sizes (varying how many documents to sample, not which documents to sample or through what selection mechanism), and random sampling is less applicable to settings where many documents may not be relevant to the research question of interest. Some recent work has incorporated retrieval into topic modeling: \citet{spielberger2025agentic} propose Agentic RAG for exploratory qualitative research, using dense vector search to iteratively retrieve documents while an LLM agent refines topic generation, but retrieval serves as a fixed architectural component rather than an evaluated design choice. Similarly, \citet{shi2024ragtag} introduce RAGtag for dynamic topic assignment, but  do not evaluate how retrieval strategy affects discovered topics.

\paragraph{Data selection in other contexts}
Several established research areas address document selection but optimize for fundamentally different objectives than exploratory topic discovery. Active learning methods strategically select samples to maximize classifier performance with minimal annotation cost \citep{settles2009active, ash2020deep, sener2018active}, evaluating success through classification accuracy and designing selection methods around fixed ground-truth label spaces, which do not exist in open-ended analyses.
Active learning has been combined with topic modeling to aid in content analysis, but these approaches still involve running an initial topic model across the corpus \citep{poursabzi-sangdeh-etal-2016-alto,li-etal-2024-improving-tenor}, or require recurring input from the user \citep{li-etal-2025-llm-struggle}.
Beyond active learning, other sampling paradigms optimize for different objectives. Coreset methods \citep{mirzasoleiman2020coresets} accelerate supervised training through gradient approximation; Determinantal Point Processes \citep{kulesza2012determinantal} and submodular maximization \citep{lin-bilmes-2011-class} enable diverse selection under predefined kernels; information retrieval diversification \citep{carbonell1998use, santos2010explicit} balances relevance and novelty for known queries. These methods assume fixed objectives (classification accuracy, computational efficiency, or search satisfaction) evaluated against known ground truth or user feedback. Exploratory corpus analyses instead seeks to uncover thematic structures where evaluation criteria emerge from domain analysis rather than predetermined categories.

\section{Methodology}
\label{sec:methodology}
\subsection{Overview}
We systematically model and compare document selection strategies in settings where the goal is to analyze documents relevant to a particular query out of larger corpus.
Consider for example, studying, ``How has the COVID-19 pandemic impacted violence in society, including violent crimes?'' from an archive of papers about COVID-19. While answering this query requires synthesizing many (100s-1000s) of documents, the majority of documents in the corpus focus about other aspects of COVID-19 and are not relevant for this query.

We design an evaluation framework to assess how document selection strategy impacts discovered topics, assuming a fixed analysis budget such as a maximum number of documents. Strategies ignoring the query produce irrelevant findings, while over-optimizing for relevance may yield near-duplicate documents with low diversity and limited coverage.
Even when query-focused, open-ended text analysis differs fundamentally from information retrieval: IR prioritizes top-ranked documents for specific information needs, while open-ended analysis requires analyzing hundreds or thousands of documents from large corpora to discover comprehensive thematic structures.

Our overall procedure is as follows: given a corpus of documents $D$, a query $q$, and a selection strategy $\mathcal{S}$, we apply $\mathcal{S}$ to $q$ and $D$ to obtain $d_{analysis} \subset D$. We fix the size of $d_{analysis}$ across all selection strategies evaluated. We then run an text analysis method over $d_{analysis}$ to obtain a set of topics $T_q = \{t_1, \dots, t_n\}$. We evaluate the relevance, diversity and coverage of $T_q$ under multiple analysis methods to assess $\mathcal{S}$, as detailed below.

\subsection{Exploratory Analysis Methods}
We use four topic models as the downstream analysis methods.
As a statistical method, we use LDA \citep{blei2003latent}, which remains a popular go-to approach for exploratory text analysis.
We also use BERTopic \citep{grootendorst2022bertopic}, which relies on clustering semantic embeddings, and two approaches based on LLM-prompting: TopicGPT \citep{pham-etal-2024-topicgpt} and HiCode \citep{zhong-etal-2025-hicode}. While TopicGPT aims to produce LDA-style topics, HiCode draws inspiration from inductive qualitative coding procedures.
For each method, we map each generated topic $t_i$ to a semantic embedding (\texttt{all-mpnet-base-v2} embeddings for the concatenated top-10 words for BERTopic/LDA, natural language descriptions for TopicGPT, and fine-grained themes for HiCode), which we use to compute evaluation metrics.
We provide implementation details in \Cref{app:topic_models}.

\subsection{Evaluation Framework}\label{sec: eval framework}
We evaluate the quality of produced topics for (1) relevancy: does a selection strategy effectively identify documents that produce relevant topics for the query? (2) diversity: how diverse are the topics in the selected data? and (3) coverage: does the set of topics sufficiently recover all of the information relevant for the query? For (3), in the absence of ground-truth knowledge of the total set of relevant information, we focus on relative coverage: did this strategy miss topics that would have been identified by a different selection strategy?

\paragraph{Topic Relevancy to Query} For relevance, we measure the cosine similarity between a semantic embedding of the query and the embedding representation for each topic. In all formulas below, $similarity(q, t)$ denotes the cosine similarity between the embeddings of query $q$ and topic $t$. We average across topics and queries:
\[relevancy(\mathcal{S}) = \dfrac{1}{|Q|}\sum_{q\in Q} \dfrac{\sum_{t \in T_{q}} similarity(q, t)}{|T_{q}|}\]

\paragraph{Diversity of Topics}
For a strategy $\mathcal{S}$, the diversity score is calculated as the average distance between each topic embedding:
\[diversity(\mathcal{S}) = \frac{1}{{|Q|}}\sum_{q\in Q} \frac{\sum_{i<j} dist(t_i, t_j)}{\binom{|T|}{2}}\] %|T| (|T| - 1)
where we use $1 - \text{cosine similarity}(t_i, t_j)$ for $dist(t_i, t_j)$. We compute overall diversity on $T_q$ and relevant diversity on relevant topics $T_q^{rel} = \{t \in T_{q} : similarity(q, t) \geq 0.5\}$, excluding noisy topics that may inflate diversity.

\paragraph{Relevant Topic Coverage between Selection Strategies}
Lastly, since we lack gold labels, we measure complementarity between selection strategies: does method $\mathcal{S}_A$ sufficiently cover all the relevant topics identified by $\mathcal{S}_B$? We define coverage ($\mathcal{S}_A \rightarrow \mathcal{S}_B$) as the fraction of $\mathcal{S}_B$'s query-relevant topics that match topics in $\mathcal{S}_A$. Low coverage indicates that $\mathcal{S}_A$ fails to identify topics identified by $\mathcal{S}_B$, and we would have found more content had we run $\mathcal{S}_B$ instead; high coverage indicates that $\mathcal{S}_A$ is sufficient. Coverage is averaged across queries:

\[coverage(\mathcal{S}_{A}, \mathcal{S}_{B}) = \dfrac{1}{|Q|}\sum_{q\in Q}\dfrac{|T_{q}^{A,rel} \bigcap T_{q}^{B,rel}|}{|T_{q}^{B,rel}|}\]
where $T_{q}^{\mathcal{S}, rel}$ are the relevant topics for the selection $\mathcal{S}$ same as in relevance and diversity scores. To empirically get the matched topics $T_{q}^{A,rel} \bigcap T_{q}^{B,rel}$, we establish the similarity matrix from $similarity(T_q^{A, rel}, T_q^{B, rel}) \geq 0.7$ and use Hungarian algorithm~\citep{kuhn1955hungarian} for optimal 1-1 topic matching. We use threshold 0.7 (selected by qualitative inspection at $\{0.5, 0.6, 0.7\}$) as matched topics should be highly similar.

\section{Experiments}
\label{sec:experiments}

\subsection{Data}
\textbf{TREC-COVID} We use the TREC-COVID test collection \citep{voorhees2020trec}, containing $171,332$ biomedical research articles with titles and abstracts and $50$ queries with gold relevance judgments for each article QREL $\in \{0,1,2\}$: irrelevant, slightly relevant, highly relevant. Prior to any experiments, we manually selected $15$ queries (App.~\ref{app:treccovid_queries}) that we expect to require synthesizing topics across many documents, rather than being answerable from 1-2 documents.
We choose queries to span a range of content, e.g. Query 43: transmission mechanisms, Query 10: chloroquine effectiveness.
As TREC-COVID is a commonly used benchmark for information retrieval, this dataset allows us to verify the implementation of our retrieval data selection strategies.

\paragraph{Doctor-Reviews}
We use a corpus of $>4$M online physician reviews originating from Healthgrades, Vitals, RateMDs, and Yelp  \citep{luo2025mapping}. For efficiency, we focus on reviews of family medicine physicians, discarding ones with abnormally many or few reviews for a final subset of $171,216$ reviews for $5,986$ physicians. Unlike TREC-COVID, this data does not have pre-curated queries and thus reflects a more real-world use case.  We consulted with public health experts to create $11$ exploratory queries that they find valuable to study (App.~\ref{app:doctor_queries_list}), such as understanding patient experiences, pathways to healthcare providers, and relationship with physicians.

\subsection{Data Selection Strategies}

We begin with ten document selection strategies representing a spectrum of modern retrieval approaches, from simple baselines to advanced semantic methods, and ultimately run full evaluations with seven, based on initial retrieval performance.
We briefly describe methods here and provide further details in  App.~\ref{app:sampling}. For all methods, we select 1,000 documents per query.

We employ three zero or single feature approaches: \textbf{Random Uniform} randomly samples documents from the full corpus without targeting the query.
\textbf{Keyword Search (BM25)} uses traditional lexical matching \citep{robertson1994okapi}. \textbf{SBERT} employs dense semantic embeddings \citep{reimers-gurevych-2019-sentence}, capturing meaning beyond exact word matches.

We employ three approaches that combine keyword and semantic features (hybrid search): Hybrid Simple Sum (which we generally refer to as \textbf{Direct Retrieval}) combines lexical and semantic signals, equally balancing keyword precision with semantic recall. \textbf{Hybrid Reciprocal Rank Fusion (RRF)} uses rank-based feature combinations \citep{cormack2009reciprocal} and \textbf{Hybrid Weighted (Wt)} uses score-based combinations \citep{fox1994combination}.

Finally we evaluate four extensions to hybrid search methods: \textbf{Hybrid + Cross-Encoder (CE)} applies neural reranking \citep{nogueira2019passage} to improve precision. \textbf{Hybrid + MMR} enforces diversity through Maximal Marginal Relevance \citep{carbonell1998use}, explicitly trading relevance for coverage. \textbf{Query Expansion with KeyBERT (QryExp)} reformulates queries by extracting keywords \citep{grootendorst2020keybert}, expanding retrieval scope.
\textbf{Retrieval Random} samples $20\%$ of documents retrieved by Direct Retrieval, simulating selection from a curated collection and aiming to improve diversity.

\begin{table*}[t]
\centering
\small
\resizebox{0.95\linewidth}{!}{
\begin{tabular}{lcccccccc}
\toprule
\multirow{2}{*}{Method} & \multicolumn{4}{c}{TREC-COVID} & \multicolumn{4}{c}{Doctor-Reviews} \\
\cmidrule(lr){2-5} \cmidrule(lr){6-9}
& TopicGPT & BERTopic & LDA & HiCode & TopicGPT & BERTopic & LDA & HiCode \\
\midrule
Random Uniform & .156$_{\pm.035}$ & .232$_{\pm.030}$ & .266$_{\pm.043}$ & .648$_{\pm.115}$ & .224$_{\pm.077}$ & .293$_{\pm.079}$ & .397$_{\pm.106}$ & .580$_{\pm.051}$ \\
Keyword Search & .326$_{\pm.072}$ & .325$_{\pm.066}$ & .575$_{\pm.082}$ & .679$_{\pm.105}$ & .341$_{\pm.106}$ & .404$_{\pm.071}$ & .543$_{\pm.082}$ & \textbf{.628}$_{\pm.033}$ \\
SBERT & .398$_{\pm.078}$ & \textbf{.458}$_{\pm.028}$ & .596$_{\pm.065}$ & \textbf{.704}$_{\pm.083}$ & \textbf{.402}$_{\pm.097}$ & .446$_{\pm.055}$ & .501$_{\pm.081}$ & .627$_{\pm.044}$ \\
Direct Retrieval & \textbf{.435}$_{\pm.135}$ & .407$_{\pm.038}$ & .606$_{\pm.071}$ & .676$_{\pm.092}$ & .324$_{\pm.116}$ & \textbf{.471}$_{\pm.120}$ & \textbf{.577}$_{\pm.076}$ & .624$_{\pm.042}$ \\
Dir. Retr. + MMR & .265$_{\pm.065}$ & .436$_{\pm.119}$ & .594$_{\pm.044}$ & .680$_{\pm.088}$ & .288$_{\pm.055}$ & .397$_{\pm.064}$ & .513$_{\pm.101}$ & .619$_{\pm.045}$ \\
Query Expansion & .415$_{\pm.104}$ & .402$_{\pm.042}$ & \textbf{.609}$_{\pm.070}$ & .641$_{\pm.197}$ & .380$_{\pm.095}$ & .464$_{\pm.120}$ & .576$_{\pm.069}$ & .624$_{\pm.058}$ \\
Retrieval Random & .309$_{\pm.067}$ & .374$_{\pm.039}$ & .536$_{\pm.060}$ & .674$_{\pm.100}$ & .316$_{\pm.100}$ & .405$_{\pm.087}$ & .493$_{\pm.098}$ & .603$_{\pm.059}$ \\
\bottomrule
\end{tabular}}
\caption{Average topic-query relevance scores~(mean $\pm$ std across queries) for all topic models and datasets. Bold indicates highest relevance in each column. Semantic retrieval methods consistently achieve higher relevance than lexical methods across TopicGPT, BERTopic, and (to a lesser extent) LDA. HiCode shows compressed ranges due to query-aware topic generation. Doctor-Reviews shows weaker differentiation and higher variance.}
\label{tab:alignment_all_models}
\end{table*}

\section{Results}
\label{sec:results}

\begin{figure}[t]
    \centering
    \includegraphics[width=1\linewidth]{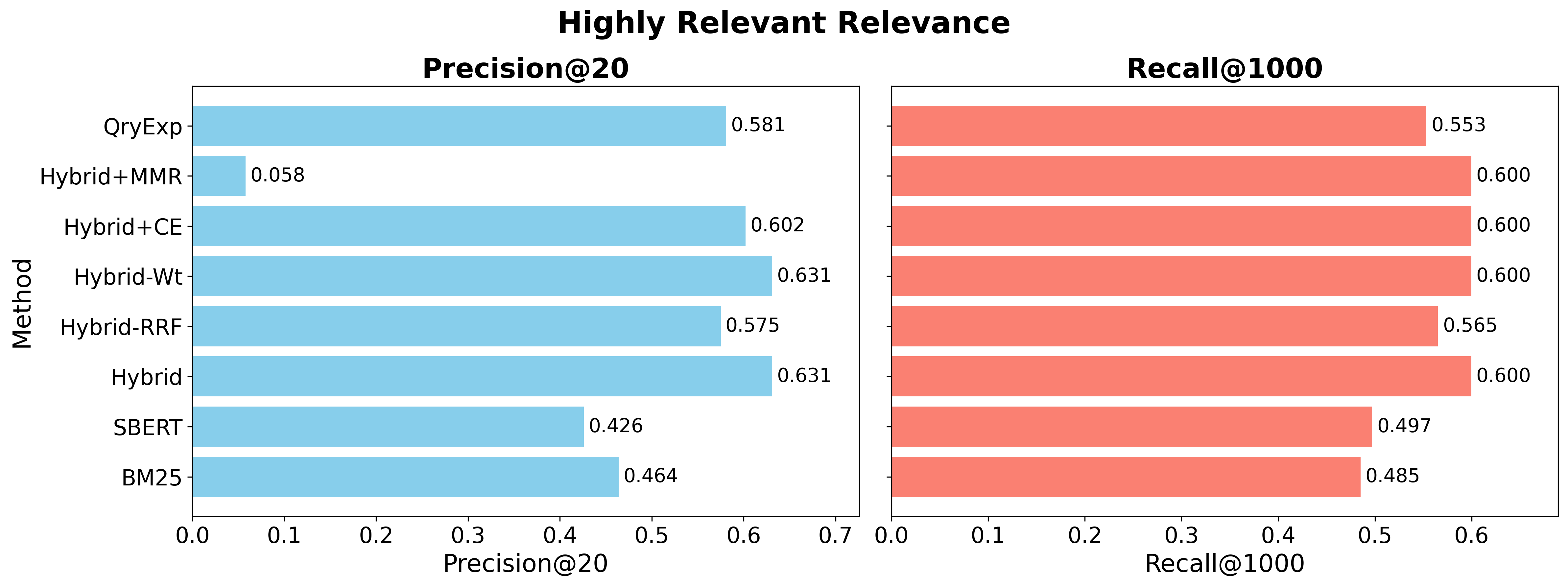}
    \caption{Information retrieval metrics across all 50 TREC-COVID queries showing precision@20 and recall@1,000. Hybrid fusion methods achieve the highest precision, while MMR diversity reranking substantially reduces precision despite maintaining recall.}
    \label{fig:irmetrics}
\end{figure}

\subsection{Retrieval Validation}
\label{sec:retrieval_validation}
We first validate our retrieval methods using the documents annotated as relevant for all 50 queries in the TREC-COVID data set (Fig.~\ref{fig:irmetrics}).
We exclude random and retrieval random from these evaluations, as they do not directly target retrieval.
As is standard, for each method, we retrieve $1,000$ documents per query and compute precision and recall for documents annotated as highly relevant (QREL = 2).
Our results align with \citet{chen2022out}'s evaluation on the same dataset, validating our implementation.
For subsequent experiments, we use Hybrid Simple Sum as our primary hybrid retrieval method (referred to as \textbf{Direct Retrieval}), as it performs equivalently or better than alternatives. We exclude CrossEncoder reranking: it adds significant cost \citep{nogueira2019passage} with limited benefits, as our evaluation focuses on which documents are selected rather than their precise ordering.

\subsection{Topic Relevance with Queries}
\label{sec:query_alignment}

Tab.~\ref{tab:alignment_all_models} shows relevance scores across selection methods for all four topic models on both datasets. Across topic models, semantic retrieval methods (SBERT, Direct Retrieval, Query Expansion) consistently achieve better relevance than Keyword Search, which in turn exceeds Random Uniform. The ranking within semantic methods varies by model: SBERT achieves highest relevance in BERTopic and HiCode, Direct Retrieval achieves highest in TopicGPT, and Query Expansion achieves highest in LDA. Random Uniform consistently shows the lowest relevance across all models except HiCode.

The most notable difference across topic models is that HiCode achieves uniformly high relevance across all methods on TREC-COVID, with Random Uniform approaching semantic methods. This reflects HiCode's query-aware topic generation mechanism: the LLM prompts explicitly include the query text, directing all discovered topics toward query concepts regardless of selection method, e.g., even if the data selection method identifies irrelevant documents, HiCode does not include them in the generated themes. The primary difference between datasets is that Doctor-Reviews shows weaker differentiation and higher variance across selection methods. This trend is primarily driven by one query (Q3: ``What breathing problems do patients report and how are they treated?'') with near-random relevance scores across all methods (mean 0.450), likely unanswerable from this data. Excluding this outlier, the remaining 10 queries show patterns consistent with TREC-COVID (mean 0.552$_{\pm0.044}$ vs 0.543$_{\pm0.052}$ overall).

\subsection{Topic Diversity Patterns}
\label{sec:alignment_diversity}

\subsubsection{Overall Semantic Diversity}

\begin{figure}[t]
    \centering
    \includegraphics[width=0.9\linewidth]{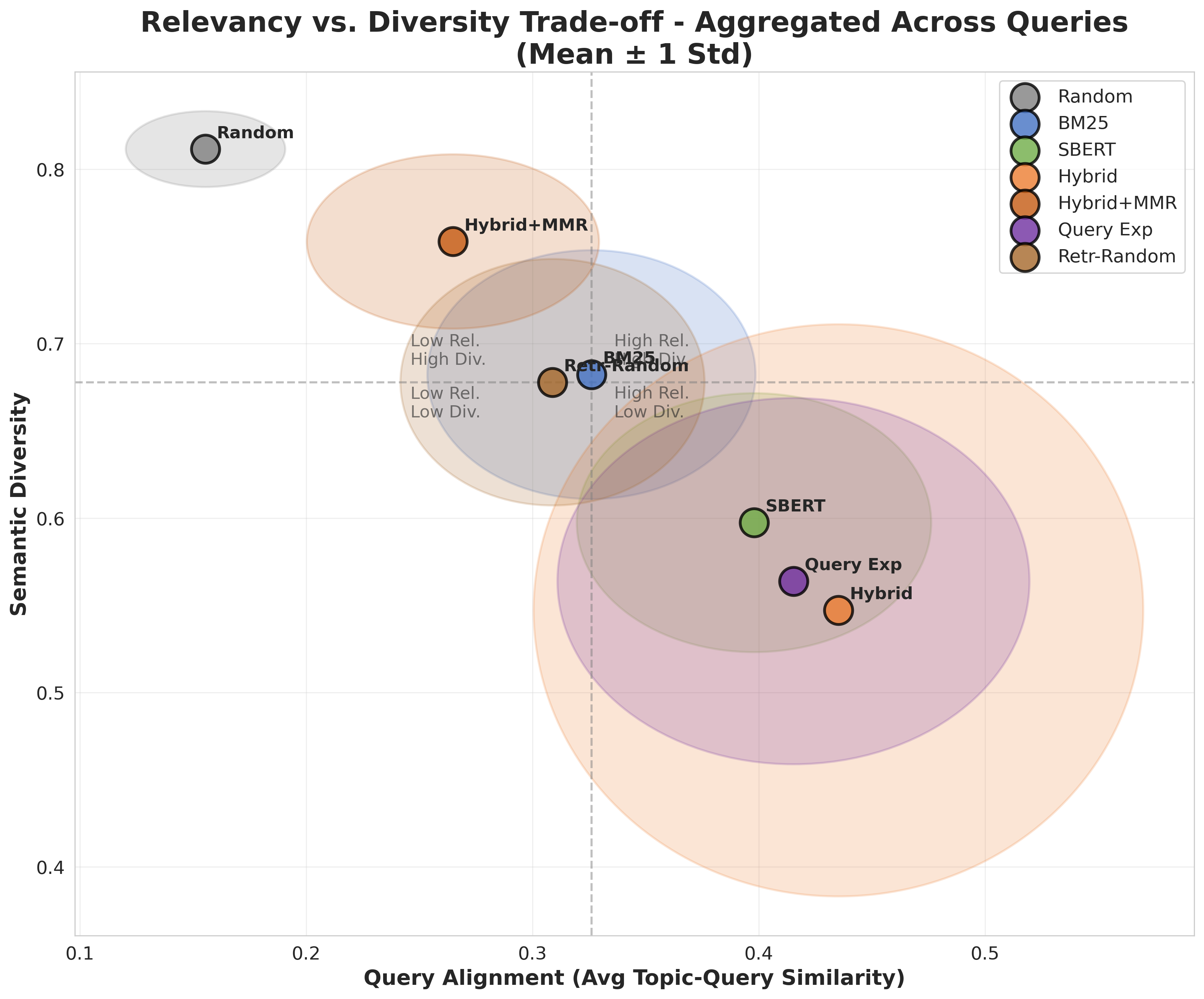}
    \caption{Topic-query relevance versus overall semantic diversity for TopicGPT on TREC-COVID. Ellipses show mean $\pm$ 1 standard deviation across 15 queries. Methods cluster along a negative relationship: higher relevance associates with lower diversity. Random Uniform occupies the low-relevance, high-diversity quadrant; Direct Retrieval, Query Expansion, and SBERT occupy the high-relevance, low-diversity quadrant.
    }
    \label{fig:relevance_diversity}
\end{figure}

Fig.~\ref{fig:relevance_diversity} plots the relevance scores against diversity scores using the full set of produced topics for TopicGPT on TREC-COVID. For brevity, results from other topic models are reported in  App.~\ref{app:diversity_all}. In this setting, relevance and diversity show an inverse relationship, suggesting that choosing a high-relevance data selection method could lower the distinctness of the identified topics, limiting the exploratory nature of the analysis.

\begin{figure*}[htb!]
    \centering
    \includegraphics[width=0.35\linewidth]{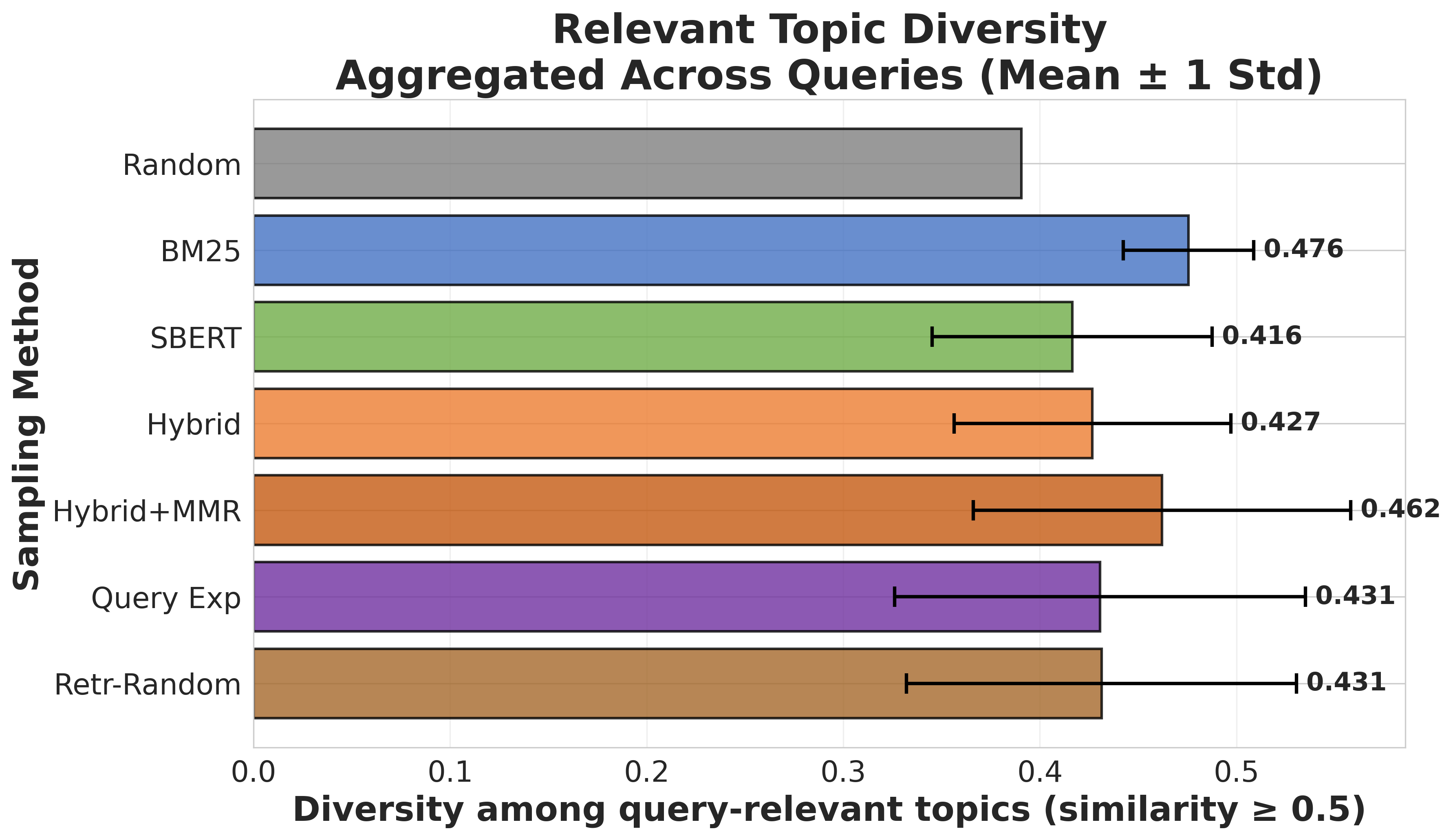}
    \includegraphics[width=0.35\linewidth]{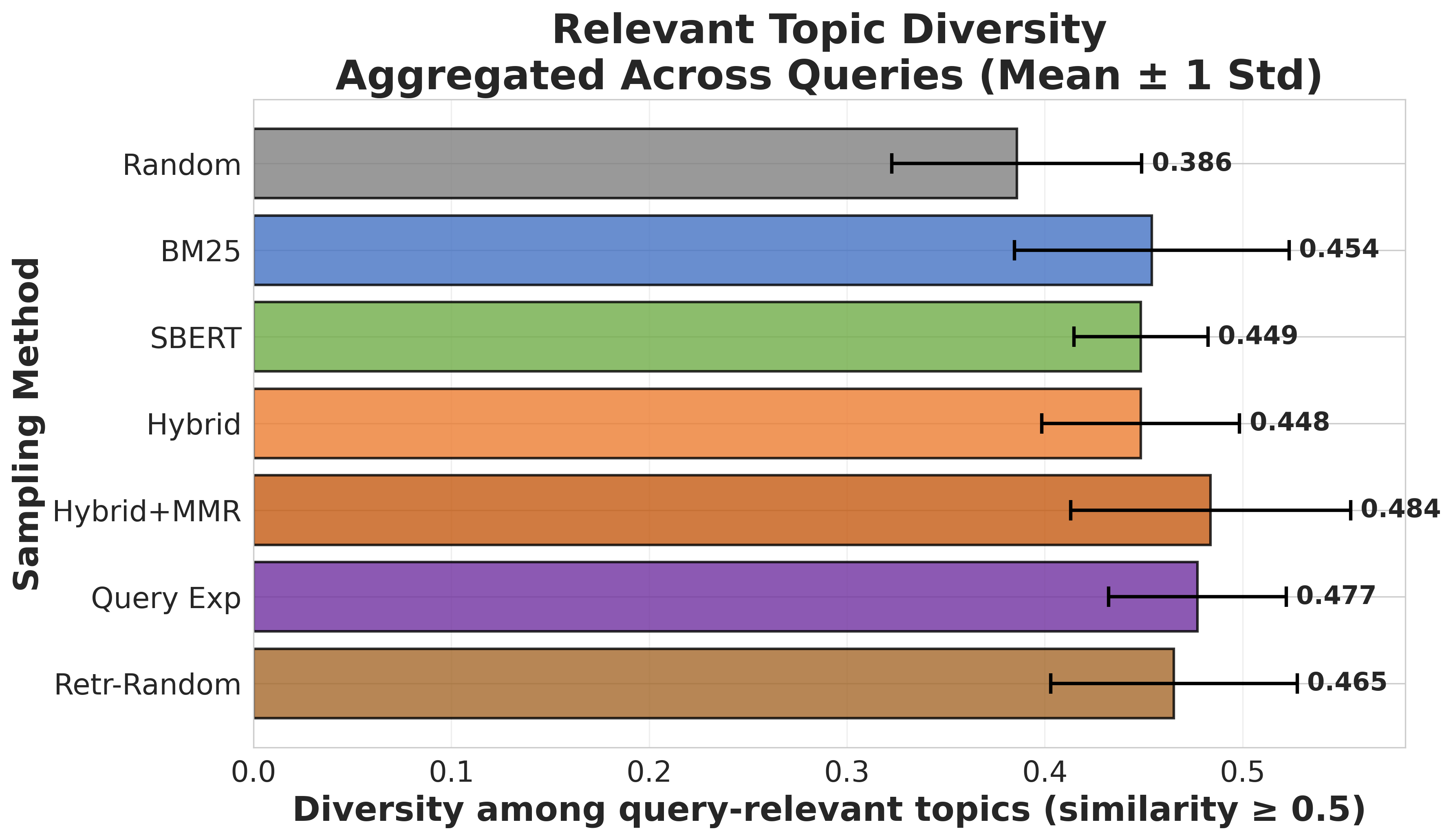}
    \caption{Semantic diversity among query-relevant topics only (similarity $>$0.5) for TopicGPT (left) and BERTopic (right) on TREC-COVID. Error bars show $\pm$1 standard deviation across 15 queries. Both models show minimal variation across selection methods, contrasting with the large differences in overall diversity in Fig.~\ref{fig:relevance_diversity}. This indicates diversity is preserved where it matters: among topics semantically related to the query.
    }
    \label{fig:relevant_diversity_trec}
\end{figure*}

This pattern appears across other topic models on TREC-COVID, though the magnitude and strictness of the relationship varies. The trade-off is most pronounced in TopicGPT and BERTopic, while LDA and HiCode show weaker negative relationships and lower diversity overall. The weakness of the trend in HiCode likely reflects the same trend in Tab.~\ref{tab:alignment_all_models}, that HiCode generates relevant topics regardless of data selection. On Doctor-Reviews (App.~\ref{app:diversity_all}), there is an inverse relationship between diversity and relevance under TopicGPT, but no noticeable trade-off in other settings. Overall these results suggest that when the intended downstream analysis method is a high-diversity model (e.g. TopicGPT), choosing a high-relevance selection strategy can undermine that diversity, but the choice of selection strategy has less impact on diversity in other settings.

\subsubsection{Diversity Among Query-Relevant Topics}

Fig.~\ref{fig:relevance_diversity} does not distinguish if additional diversity results from topics related to the research question or  random and unhelpful content. In Fig.~\ref{fig:relevant_diversity_trec}, we make this distinction by computing diversity only among relevant topics for TopicGPT and BERTopic on TREC-COVID. Unlike overall diversity, relevant topic diversity shows minimal variation across selection methods.
Random Uniform does show lower relevant diversity, but this difference likely reflects the scarcity of relevant topics discovered by random selection rather than differences in how diverse those few relevant topics are.
The remaining topic models on TREC-COVID and all topic models on Doctor-Reviews show a similar lack of differentiation between data selection methods for relevant diversity metrics (App.~\ref{app:relevant_div_all}).

This pattern of large differences in overall diversity and small differences in relevant diversity, indicates that the diversity ``lost'' through retrieval bias primarily affects off-topic or tangentially related topics rather than query-relevant topics. The high overall diversity of random selection includes substantial contribution from irrelevant topics, while retrieval methods maintain similar diversity where it matters: among topics semantically related to the information need.
Thus, while Fig.~\ref{fig:relevance_diversity} suggests that sacrificing relevance and accepting noisier results may be necessary in settings where diversity is paramount, Fig.~\ref{fig:relevant_diversity_trec} suggests that prioritizing relevance is appropriate in most settings.

\subsection{Pairwise Topic Coverage Analysis}
\label{sec:coverage}

Beyond relevance and diversity, we examine how topics discovered by different selection methods compare through pairwise relevant topic coverage, thus evaluating, would we have identified more topics if we had run method B instead of A?

\begin{figure*}[t]
    \centering
    \includegraphics[width=0.38\linewidth]{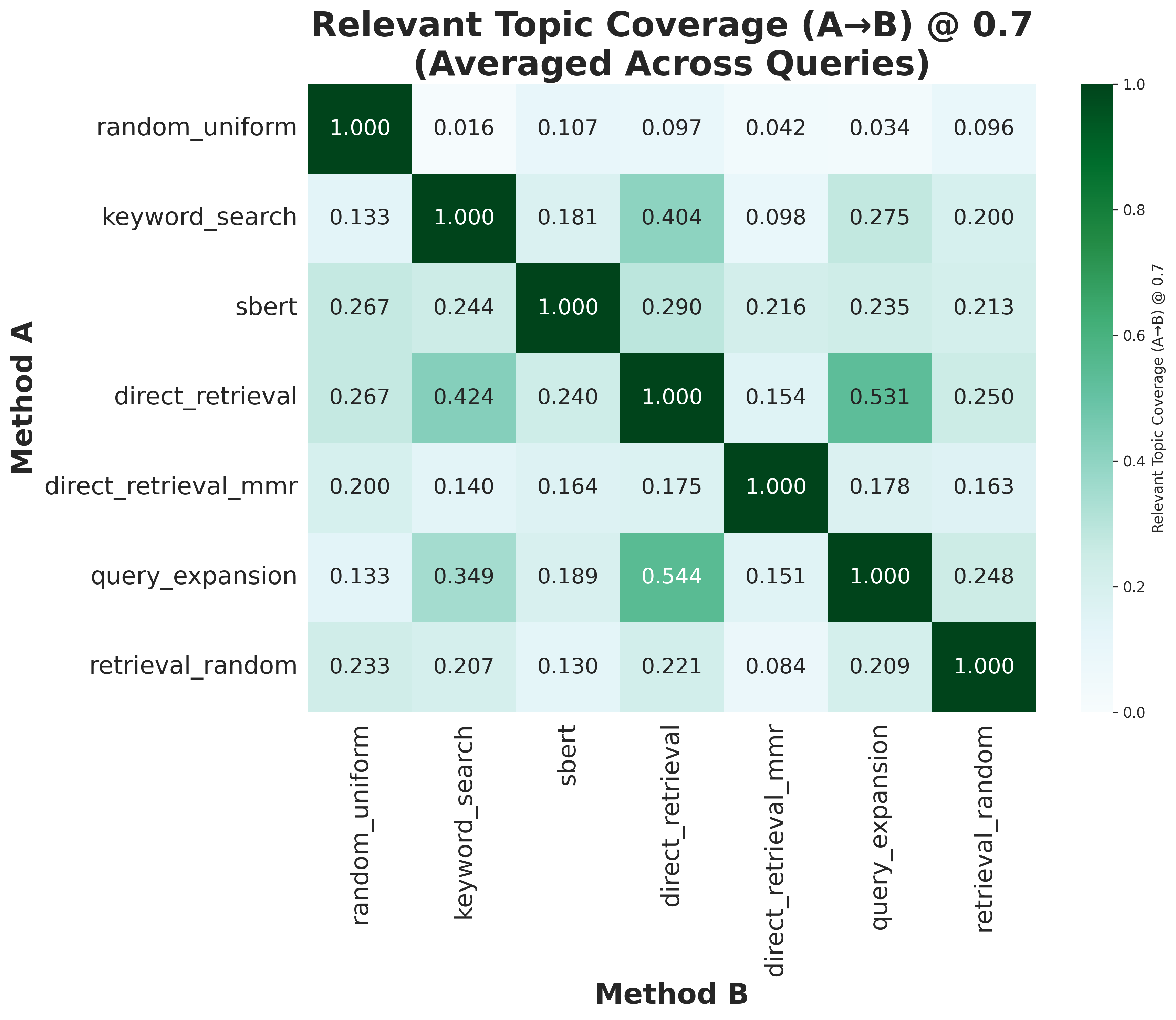}
    \includegraphics[width=0.38\linewidth]{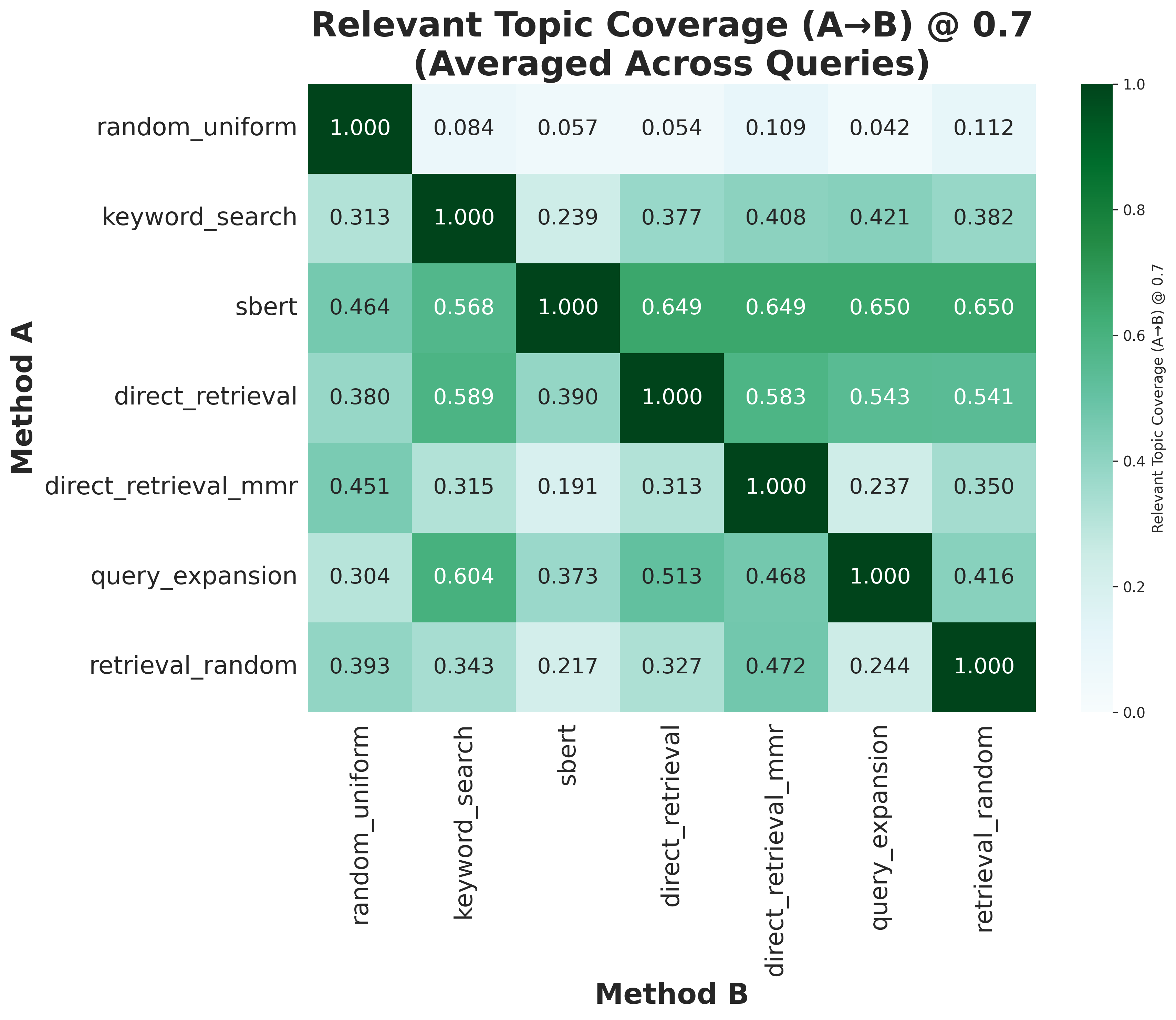}
    \caption{Pairwise relevant topic coverage at semantic similarity threshold 0.7 for TopicGPT (left) and BERTopic (right) on TREC-COVID. Cell (i,j) shows the fraction of method j's relevant topics covered by method i. \textbf{Reading guide:} Dark row = comprehensive (covers many others' topics). Light row = Low coverage. Dark column = redundant (widely covered by others). Light column = unique contribution.}
    \label{fig:coverage_trec}
\end{figure*}

Fig.~\ref{fig:coverage_trec} shows relevant topic coverage heatmaps for TopicGPT and BERTopic on TREC-COVID. Semantic retrieval methods (SBERT, Direct Retrieval, Query Expansion) show darker row patterns, indicating they converge on similar sets of query-relevant topics. In contrast, Random Uniform shows very light row patterns, covering minimal fractions of other methods' relevant topics.

This pattern appears across both clustering-based (BERTopic) and LLM-based (TopicGPT) topic modeling approaches on TREC-COVID. On Doctor-Reviews, the convergence among semantic methods and minimal coverage by Random Uniform hold consistently, though relative method strengths vary: Keyword Search shows stronger coverage than on TREC-COVID, while Direct Retrieval remains the most comprehensive across both datasets.
Appendix~\ref{app:coverage_all} provides figures for all topic models and datasets.

\subsubsection{Case Study: Violence and COVID-19}

To better illustrate coverage patterns, we examine a query from TREC-COVID using TopicGPT: \textit{``How has the COVID-19 pandemic impacted violence in society, including violent crimes?''} with complete topic lists in App.~\ref{app:query43_topics}.

Keyword Search discovered 59 topics with 5 relevant topics. Approximately 300 documents (30\% of the sample) were assigned to topics about medical specialty organizations (surgical societies, cardiology, gastroenterology) because the query term ``society'' lexically matched ``surgical society'' and ``American Society of'', despite being semantically unrelated to violence. This lexical polysemy artifact introduced substantial topic noise.
Semantic methods avoided this artifact and discovered topics absent in Keyword Search results. SBERT uniquely identified violence against health workers and violence exposure in youth as separate topics. Query Expansion discovered gun violence despite ``gun'' not appearing in the original query. These nuanced aspects were not captured by lexical retrieval.

The case study demonstrates the qualitative patterns quantified in preceding sections: semantic methods (SBERT, Direct Retrieval, Query Expansion) discover nuanced, query-relevant topics and avoid lexical artifacts, while Keyword Search captures some relevant topics alongside substantial noise from term ambiguity. SBERT's fine-grained topic separation and discovery of unique violence-related aspects exemplify the distinct topic profiles produced by different retrieval strategies.

\section{Discussion}
\label{sec:discussion}

Our evaluation across four topic models, two datasets, and 26 queries reveals systematic patterns in how selection methods affect topic discovery. 

\subsection{Beyond Classical Information Retrieval} Classical information retrieval optimizes for precision and recall, retrieving relevant documents from large collections, while topic modeling for exploratory analysis requires both relevance and diversity among discovered themes. This task difference produces divergent method rankings. SBERT and Keyword Search achieve comparable IR metrics (Fig.~\ref{fig:irmetrics}), yet differ substantially in topic-query relevance across models (Tab.~\ref{tab:alignment_all_models}). Query Expansion, traditionally effective in IR~\citep{carpineto2012survey}, provides minimal relevance improvements over Direct Retrieval. MMR diversity reranking substantially reduces IR precision (Fig.~\ref{fig:irmetrics}) but maintains reasonable topic relevance with queries, albeit inconsistently (Tab.~\ref{tab:alignment_all_models}). Our coverage and diversity metrics have no direct IR analogue. Recall measures document retrieval completeness but does not capture thematic diversity. Given these task-specific requirements, we offer practical guidelines based on our evaluation.

\subsection{General Recommendations}

\textbf{Default: SBERT or Direct Retrieval.} SBERT and Direct Retrieval offer strong overall performance compared to alternatives. Both achieve substantially higher topic-query relevance than Keyword Search.
SBERT is easier to implement, requiring only semantic embeddings without maintaining a BM25 index, but
Direct Retrieval's high relevance (Tab.~\ref{tab:alignment_all_models}) may provide robustness when consistent performance is critical, and hybrid methods support potential parallel processing of keyword and semantic search, reducing latency.

\textbf{Avoid random sampling.} Random Uniform results in fewer relevant topics than semantic methods (Tab.~\ref{tab:alignment_all_models}). While random sampling maximizes overall diversity (Fig.~\ref{fig:relevance_diversity}), diversity is driven by off-topic content (Fig.~\ref{fig:relevant_diversity_trec}).
Random sampling remains appropriate for broad corpus characterization but not work focused on a specific research question.

\textbf{Avoid Query Expansion.} Query Expansion provides minimal relevance improvements over Direct Retrieval (Tab.~\ref{tab:alignment_all_models}) despite requiring keyword extraction and fusion logic. Coverage analysis (Fig.~\ref{fig:coverage_trec}) shows substantial overlap with Direct Retrieval at threshold 0.7, indicating similar topic structures with minor reweighting. The computational overhead is not justified by marginal gains.

\textbf{MMR diversity enforcement.} MMR provides tunable relevance-diversity trade-offs but shows high variance across topic models and datasets. We recommend MMR only when diversity-relevance control is critical to analytical goals and resources permit hyperparameter tuning.

\textbf{Keyword Search: polysemy failures.} Our case study demonstrates systematic failure: ``violence in society'' matched ``surgical society'' (Sec. ~\ref{sec:coverage}).
Keyword Search should only be considered when: (1) queries use technical terminology with minimal polysemy, (2) resource constraints prohibit embeddings, and (3) manual filtering is feasible.

\subsection{Additional Considerations}
While we primarily focus on evaluating data selection strategies, our results also reveal how query and topic model design influence the effectiveness of different selection strategies.

\textbf{Query Quality}
As noted in Sec.~\ref{sec:query_alignment}, Query 3 in Doctor-Reviews (``What breathing problems do patients report and how are they treated?'') resulted in near-random relevance all models. This query combined multiple concepts (problems AND treatment), used vague terminology (``breathing problems''), and may mismatch corpus content (reviews rarely detail clinical treatments). This result serves as a reminder that the choice of methodology has little impact for a research question that is poorly formulated or unanswerable from the data.

\textbf{Topic Model Considerations}
LDA and HiCode overall demonstrated narrower differentiation of selection methods than TopicGPT and BERTopic, suggesting that the choice of data selection strategy is less impactful if the planned downstream analysis method is one of these two approaches.
Both models do still demonstrate observable differences, e.g. between Random Uniform and semantic methods (App.~\ref{app:coverage_all}), suggesting the choice of data selection strategy cannot be ignored entirely.
Notably, Keyword Search performs better with LDA than with other topic models. When computational constraints necessitate keyword-only retrieval, pairing Keyword Search with LDA may provide reasonable performance, though this conclusion requires further validation in specific domains.

\paragraph{Conclusion}
\label{sec:conclusion}
Our work introduces data selection for corpus analysis as a distinct task, not captured by classical information retrieval metrics. Through robust evaluations, we offer practice guidance: semantic or hybrid retrieval provides strong default performance, current advanced methods offer minimal improvements, and keyword-based approaches suffer from  failures on open-ended queries.
While our work focuses on evaluating existing selection strategies, our findings and evaluation framework can additionally support developing new ones.
Overall, this work formulates document sampling as a deliberate methodological choice, providing an evaluation framework that generalizes to future selection strategies and enables principled decisions based on analytical goals.

\section*{Limitations}
\label{sec:limitations}

Our evaluation covers two datasets from specific domains (biomedical literature, healthcare reviews) with some manually-designed queries, requiring validation on other domains and query types. We fix sample size at 1,000 documents to isolate selection strategy effects but do not explore how sample size interacts with retrieval method. While we evaluate four architecturally diverse topic models and seven selection strategies, many alternative approaches exist; our findings may not generalize to all modeling or retrieval methods. We rely on automatic relevance and diversity metrics without ground-truth topic labels, though qualitative case studies validate key findings. Our relevance threshold (0.5 or 0.7) is informed by qualitative analysis but may not be optimal across all settings, though results remain consistent across thresholds 0.5-0.7.

\section*{Acknowledgement}
This work is funded by the Sanofi iDEA-TECH Award program.

\bibliography{latex/custom, latex/anthology}

\begin{thebibliography}{38}
\providecommand{\natexlab}[1]{#1}

\bibitem[{Anderson et~al.(2011)Anderson, McCandless, Klausner, Taketa, and Yerger}]{anderson2011tobacco}
Stacey~J Anderson, Phyra~M McCandless, Kim Klausner, Rachel Taketa, and Valerie~B Yerger. 2011.
\newblock \href {https://doi.org/10.1136/tc.2010.041921} {Tobacco documents research methodology}.
\newblock \emph{Tobacco Control}, 20(Suppl 2):ii8--ii11.

\bibitem[{Ash et~al.(2020)Ash, Zhang, Krishnamurthy, Langford, and Agarwal}]{ash2020deep}
Jordan~T. Ash, Chicheng Zhang, Akshay Krishnamurthy, John Langford, and Alekh Agarwal. 2020.
\newblock \href {https://openreview.net/forum?id=ryghZJBKPS} {Deep batch active learning by diverse, uncertain gradient lower bounds}.
\newblock In \emph{International Conference on Learning Representations}.

\bibitem[{Blei et~al.(2003)Blei, Ng, and Jordan}]{blei2003latent}
David~M. Blei, Andrew~Y. Ng, and Michael~I. Jordan. 2003.
\newblock \href {https://dl.acm.org/doi/10.5555/944919.944937} {Latent dirichlet allocation}.
\newblock \emph{J. Mach. Learn. Res.}, 3(null):993–1022.

\bibitem[{Caleb~Alexander et~al.(2022)Caleb~Alexander, Mix, Choudhury, Taketa, Tomori, Mooghali, Fan, Mars, Ciccarone, Patton, Apollonio, Schmidt, Steinman, Greene, Knight, Ling, Seymour, Glantz, and Tasker}]{caleb2022opioid}
G.~Caleb~Alexander, Lisa~A. Mix, Sayeed Choudhury, Rachel Taketa, Cecília Tomori, Maryam Mooghali, Anni Fan, Sarah Mars, Dan Ciccarone, Mark Patton, Dorie~E. Apollonio, Laura Schmidt, Michael~A. Steinman, Jeremy Greene, Kelly~R. Knight, Pamela~M. Ling, Anne~K. Seymour, Stanton Glantz, and Kate Tasker. 2022.
\newblock \href {https://doi.org/10.2105/AJPH.2022.306951} {The opioid industry documents archive: A living digital repository}.

\bibitem[{Carbonell and Goldstein(1998)}]{carbonell1998use}
Jaime Carbonell and Jade Goldstein. 1998.
\newblock \href {https://doi.org/10.1145/290941.291025} {The use of {MMR}, diversity-based reranking for reordering documents and producing summaries}.
\newblock In \emph{Proceedings of the 21st Annual International ACM SIGIR Conference on Research and Development in Information Retrieval}, SIGIR '98, page 335–336, New York, NY, USA. Association for Computing Machinery.

\bibitem[{Carpineto and Romano(2012)}]{carpineto2012survey}
Claudio Carpineto and Giovanni Romano. 2012.
\newblock \href {https://doi.org/10.1145/2071389.2071390} {A survey of automatic query expansion in information retrieval}.
\newblock \emph{ACM Comput. Surv.}, 44(1).

\bibitem[{Chen et~al.(2022)Chen, Zhang, Lu, Bendersky, and Najork}]{chen2022out}
Tao Chen, Mingyang Zhang, Jing Lu, Michael Bendersky, and Marc Najork. 2022.
\newblock \href {https://doi.org/10.1007/978-3-030-99736-6_7} {Out-of-domain semantics to the rescue! {Z}ero-shot hybrid retrieval models}.
\newblock In \emph{Advances in Information Retrieval: 44th European Conference on IR Research, ECIR 2022, Stavanger, Norway, April 10–14, 2022, Proceedings, Part I}, page 95–110, Berlin, Heidelberg. Springer-Verlag.

\bibitem[{Cormack et~al.(2009)Cormack, Clarke, and Buettcher}]{cormack2009reciprocal}
Gordon~V. Cormack, Charles L~A Clarke, and Stefan Buettcher. 2009.
\newblock \href {https://doi.org/10.1145/1571941.1572114} {Reciprocal rank fusion outperforms condorcet and individual rank learning methods}.
\newblock In \emph{Proceedings of the 32nd International ACM SIGIR Conference on Research and Development in Information Retrieval}, SIGIR '09, page 758–759, New York, NY, USA. Association for Computing Machinery.

\bibitem[{Eisenkraft~Klein et~al.(2024)Eisenkraft~Klein, MacKenzie, Hawkins, and Koon}]{klein2024inside}
Daniel Eisenkraft~Klein, Ross MacKenzie, Ben Hawkins, and Adam~D. Koon. 2024.
\newblock \href {https://doi.org/10.1215/03616878-11186127} {Inside “operation change agent”: Mallinckrodt’s plan for capturing the opioid market}.
\newblock \emph{Journal of Health Politics, Policy and Law}, 49(4):599–630.

\bibitem[{Fitzpatrick et~al.(2022)Fitzpatrick, Bertscher, and Gilmore}]{fitzpatrick_2022_corporate}
Iona Fitzpatrick, Adam Bertscher, and Anna~B. Gilmore. 2022.
\newblock \href {https://doi.org/10.1371/journal.pgph.0000379} {Identifying misleading corporate narratives: The application of linguistic and qualitative methods to commercial determinants of health research}.
\newblock \emph{PLOS Global Public Health}, 2(11):1--13.

\bibitem[{Fox and Shaw(1994)}]{fox1994combination}
Edward~A Fox and Joseph~A Shaw. 1994.
\newblock Combination of multiple searches.
\newblock \emph{NIST special publication SP}, 243.

\bibitem[{Grootendorst(2020)}]{grootendorst2020keybert}
Maarten Grootendorst. 2020.
\newblock \href {https://doi.org/10.5281/zenodo.4461265} {{KeyBERT}: Minimal keyword extraction with {BERT}.}

\bibitem[{Grootendorst(2022)}]{grootendorst2022bertopic}
Maarten Grootendorst. 2022.
\newblock \href {https://arxiv.org/abs/2203.05794} {{BERTopic}: Neural topic modeling with a class-based tf-idf procedure}.
\newblock \emph{Preprint}, arXiv:2203.05794.

\bibitem[{Kuhn(1955)}]{kuhn1955hungarian}
H.~W. Kuhn. 1955.
\newblock \href {https://doi.org/10.1002/nav.3800020109} {The {H}ungarian method for the assignment problem}.
\newblock \emph{Naval Research Logistics Quarterly}, 2(1-2):83--97.

\bibitem[{Kulesza and Taskar(2012)}]{kulesza2012determinantal}
Alex Kulesza and Ben Taskar. 2012.
\newblock \href {https://dl.acm.org/doi/10.5555/2481023} {\emph{Determinantal Point Processes for Machine Learning}}.
\newblock Now Publishers Inc., Hanover, MA, USA.

\bibitem[{Lam et~al.(2024)Lam, Teoh, Landay, Heer, and Bernstein}]{lam2024lloom}
Michelle~S. Lam, Janice Teoh, James~A. Landay, Jeffrey Heer, and Michael~S. Bernstein. 2024.
\newblock \href {https://doi.org/10.1145/3613904.3642830} {Concept induction: Analyzing unstructured text with high-level concepts using {LLooM}}.
\newblock In \emph{Proceedings of the 2024 CHI Conference on Human Factors in Computing Systems}, CHI '24, New York, NY, USA. Association for Computing Machinery.

\bibitem[{Li et~al.(2025)Li, Calvo-Bartolom{\'e}, Hoyle, Xu, Stephens, Dima, Fung, and Boyd-Graber}]{li-etal-2025-llm-struggle}
Zongxia Li, Lorena Calvo-Bartolom{\'e}, Alexander Hoyle, Paiheng Xu, Daniel Stephens, Alden Dima, Juan~Francisco Fung, and Jordan Boyd-Graber. 2025.
\newblock \href {https://doi.org/10.18653/v1/2025.acl-long.375} {Large language models struggle to describe the haystack without human help: A social science-inspired evaluation of topic models}.
\newblock In \emph{Proceedings of the 63rd Annual Meeting of the Association for Computational Linguistics (Volume 1: Long Papers)}, pages 7583--7604, Vienna, Austria. Association for Computational Linguistics.

\bibitem[{Li et~al.(2024)Li, Mao, Stephens, Goel, Walpole, Dima, Fung, and Boyd-Graber}]{li-etal-2024-improving-tenor}
Zongxia Li, Andrew Mao, Daniel Stephens, Pranav Goel, Emily Walpole, Alden Dima, Juan Fung, and Jordan Boyd-Graber. 2024.
\newblock \href {https://doi.org/10.18653/v1/2024.eacl-long.51} {Improving the {TENOR} of labeling: Re-evaluating topic models for content analysis}.
\newblock In \emph{Proceedings of the 18th Conference of the European Chapter of the Association for Computational Linguistics (Volume 1: Long Papers)}, pages 840--859, St. Julian{'}s, Malta. Association for Computational Linguistics.

\bibitem[{Lin and Bilmes(2011)}]{lin-bilmes-2011-class}
Hui Lin and Jeff Bilmes. 2011.
\newblock \href {https://aclanthology.org/P11-1052/} {A class of submodular functions for document summarization}.
\newblock In \emph{Proceedings of the 49th Annual Meeting of the Association for Computational Linguistics: Human Language Technologies}, pages 510--520, Portland, Oregon, USA. Association for Computational Linguistics.

\bibitem[{Luo et~al.(2025)Luo, Han, Welivita, Di, Wu, Zhi, Agarwal, and Gao}]{luo2025mapping}
Junjie Luo, Rui Han, Arshana Welivita, Zeleikun Di, Jingfu Wu, Xuzhe Zhi, Ritu Agarwal, and Gordon Gao. 2025.
\newblock \href {https://arxiv.org/abs/2510.03997} {Mapping patient-perceived physician traits from nationwide online reviews with {LLM}s}.
\newblock \emph{Preprint}, arXiv:2510.03997.

\bibitem[{Maier et~al.(2020)Maier, Niekler, Wiedemann, and Stoltenberg}]{maier2020document}
Daniel Maier, Andreas Niekler, Gregor Wiedemann, and Daniela Stoltenberg. 2020.
\newblock \href {https://doi.org/10.5117/CCR2020.2.001.MAIE} {How document sampling and vocabulary pruning affect the results of topic models}.
\newblock \emph{Computational Communication Research}, 2(2):139–152.

\bibitem[{Mirzasoleiman et~al.(2020)Mirzasoleiman, Bilmes, and Leskovec}]{mirzasoleiman2020coresets}
Baharan Mirzasoleiman, Jeff Bilmes, and Jure Leskovec. 2020.
\newblock \href {https://proceedings.mlr.press/v119/mirzasoleiman20a.html} {Coresets for data-efficient training of machine learning models}.
\newblock In \emph{Proceedings of the 37th International Conference on Machine Learning}, volume 119 of \emph{Proceedings of Machine Learning Research}, pages 6950--6960. PMLR.

\bibitem[{Nogueira and Cho(2020)}]{nogueira2019passage}
Rodrigo Nogueira and Kyunghyun Cho. 2020.
\newblock \href {https://arxiv.org/abs/1901.04085} {Passage re-ranking with {BERT}}.
\newblock \emph{Preprint}, arXiv:1901.04085.

\bibitem[{Pham et~al.(2024)Pham, Hoyle, Sun, Resnik, and Iyyer}]{pham-etal-2024-topicgpt}
Chau~Minh Pham, Alexander Hoyle, Simeng Sun, Philip Resnik, and Mohit Iyyer. 2024.
\newblock \href {https://doi.org/10.18653/v1/2024.naacl-long.164} {{T}opic{GPT}: A prompt-based topic modeling framework}.
\newblock In \emph{Proceedings of the 2024 Conference of the North American Chapter of the Association for Computational Linguistics: Human Language Technologies (Volume 1: Long Papers)}, pages 2956--2984, Mexico City, Mexico. Association for Computational Linguistics.

\bibitem[{Poursabzi-Sangdeh et~al.(2016)Poursabzi-Sangdeh, Boyd-Graber, Findlater, and Seppi}]{poursabzi-sangdeh-etal-2016-alto}
Forough Poursabzi-Sangdeh, Jordan Boyd-Graber, Leah Findlater, and Kevin Seppi. 2016.
\newblock \href {https://doi.org/10.18653/v1/P16-1110} {{ALTO}: Active learning with topic overviews for speeding label induction and document labeling}.
\newblock In \emph{Proceedings of the 54th Annual Meeting of the Association for Computational Linguistics (Volume 1: Long Papers)}, pages 1158--1169, Berlin, Germany. Association for Computational Linguistics.

\bibitem[{{\v R}eh{\r u}{\v r}ek and Sojka(2010)}]{vrehuuvrek2010software}
Radim {\v R}eh{\r u}{\v r}ek and Petr Sojka. 2010.
\newblock Software framework for topic modelling with large corpora.
\newblock In \emph{{Proceedings of the LREC 2010 Workshop on New Challenges for NLP Frameworks}}, pages 45--50, Valletta, Malta. ELRA.
\newblock \url{http://is.muni.cz/publication/884893/en}.

\bibitem[{Reimers and Gurevych(2019)}]{reimers-gurevych-2019-sentence}
Nils Reimers and Iryna Gurevych. 2019.
\newblock \href {https://doi.org/10.18653/v1/D19-1410} {Sentence-{BERT}: Sentence embeddings using {S}iamese {BERT}-networks}.
\newblock In \emph{Proceedings of the 2019 Conference on Empirical Methods in Natural Language Processing and the 9th International Joint Conference on Natural Language Processing (EMNLP-IJCNLP)}, pages 3982--3992, Hong Kong, China. Association for Computational Linguistics.

\bibitem[{Roberts et~al.(2014)Roberts, Stewart, Tingley, Lucas, Leder-Luis, Gadarian, Albertson, and Rand}]{roberts2014structural}
Margaret~E. Roberts, Brandon~M. Stewart, Dustin Tingley, Christopher Lucas, Jetson Leder-Luis, Shana~Kushner Gadarian, Bethany Albertson, and David~G. Rand. 2014.
\newblock \href {https://doi.org/10.1111/ajps.12103} {Structural topic models for open-ended survey responses}.
\newblock \emph{American Journal of Political Science}, 58(4):1064--1082.

\bibitem[{Robertson et~al.(1994)Robertson, Walker, Jones, Hancock{-}Beaulieu, and Gatford}]{robertson1994okapi}
Stephen~E. Robertson, Steve Walker, Susan Jones, Micheline Hancock{-}Beaulieu, and Mike Gatford. 1994.
\newblock \href {http://trec.nist.gov/pubs/trec3/papers/city.ps.gz} {Okapi at {TREC-3}}.
\newblock In \emph{Proceedings of The Third Text REtrieval Conference, {TREC} 1994, Gaithersburg, Maryland, USA, November 2-4, 1994}, volume 500-225 of \emph{{NIST} Special Publication}, pages 109--126. National Institute of Standards and Technology {(NIST)}.

\bibitem[{Santos et~al.(2010)Santos, Peng, Macdonald, and Ounis}]{santos2010explicit}
Rodrygo L.~T. Santos, Jie Peng, Craig Macdonald, and Iadh Ounis. 2010.
\newblock \href {https://doi.org/10.1007/978-3-642-12275-0_11} {Explicit search result diversification through sub-queries}.
\newblock ECIR'2010, page 87–99, Berlin, Heidelberg. Springer-Verlag.

\bibitem[{Sener and Savarese(2018)}]{sener2018active}
Ozan Sener and Silvio Savarese. 2018.
\newblock \href {https://openreview.net/forum?id=H1aIuk-RW} {Active learning for convolutional neural networks: A core-set approach}.
\newblock In \emph{International Conference on Learning Representations}.

\bibitem[{Settles(2009)}]{settles2009active}
Burr Settles. 2009.
\newblock \href {https://research.cs.wisc.edu/techreports/2009/TR1648.pdf} {Active learning literature survey}.
\newblock Technical report, University of Wisconsin-Madison Department of Computer Sciences.

\bibitem[{Shi et~al.(2025)Shi, Nie, Ning, Li, Liu, and Huo}]{shi2024ragtag}
Zhuofan Shi, Shuyang Nie, Shiqi Ning, Fan Li, Kun Liu, and Tong Huo. 2025.
\newblock \href {https://link.springer.com/chapter/10.1007/978-981-96-7175-5_5} {Ragtag: A retrieval-augmented generation-based topic modeling framework}.
\newblock In \emph{Data Mining and Big Data}, pages 57--68, Singapore. Springer Nature Singapore.

\bibitem[{Spielberger et~al.(2025)Spielberger, Artinger, Reb, and Kerschreiter}]{spielberger2025agentic}
Gerion Spielberger, Florian~M. Artinger, Jochen Reb, and Rudolf Kerschreiter. 2025.
\newblock \href {https://arxiv.org/abs/2502.20963} {Retrieval augmented generation for topic modeling in organizational research: An introduction with empirical demonstration}.
\newblock \emph{Preprint}, arXiv:2502.20963.

\bibitem[{Srivastava and Sutton(2017)}]{srivastava2017autoencoding}
Akash Srivastava and Charles Sutton. 2017.
\newblock \href {https://openreview.net/forum?id=BybtVK9lg} {Autoencoding variational inference for topic models}.
\newblock In \emph{International Conference on Learning Representations}.

\bibitem[{Voorhees et~al.(2021)Voorhees, Alam, Bedrick, Demner-Fushman, Hersh, Lo, Roberts, Soboroff, and Wang}]{voorhees2020trec}
Ellen Voorhees, Tasmeer Alam, Steven Bedrick, Dina Demner-Fushman, William~R. Hersh, Kyle Lo, Kirk Roberts, Ian Soboroff, and Lucy~Lu Wang. 2021.
\newblock \href {https://doi.org/10.1145/3451964.3451965} {{TREC-COVID}: constructing a pandemic information retrieval test collection}.
\newblock \emph{SIGIR Forum}, 54(1).

\bibitem[{Yakubi et~al.(2022)Yakubi, Gac, and Apollonio}]{Yakubi_Gac_Apollonio_2022}
Hanna Yakubi, Brian Gac, and Dorie~E. Apollonio. 2022.
\newblock \href {https://doi.org/10.1136/bmjopen-2021-052636} {Industry strategies to market opioids to children and women in the usa: a content analysis of internal industry documents from 1999 to 2017 released in state of oklahoma v. purdue pharma, l.p. et al}.

\bibitem[{Zhong et~al.(2025)Zhong, Wang, and Field}]{zhong-etal-2025-hicode}
Mian Zhong, Pristina Wang, and Anjalie Field. 2025.
\newblock \href {https://doi.org/10.18653/v1/2025.emnlp-main.1580} {{HIC}ode: Hierarchical inductive coding with {LLM}s}.
\newblock In \emph{Proceedings of the 2025 Conference on Empirical Methods in Natural Language Processing}, pages 31060--31078, Suzhou, China. Association for Computational Linguistics.

\end{thebibliography}

\appendix

\section{Data: Evaluation Queries}
\label{app:queries}
We provide the query text selected and used for TREC-COVID and Online Doctor Review datasets.
\subsection{TREC-COVID Queries}
\label{app:treccovid_queries}

We evaluate on 15 queries from the TREC-COVID test collection \citep{voorhees2020trec}:

\begin{enumerate}
\setcounter{enumi}{1}
\item[2.] How does the coronavirus respond to changes in the weather?
\item[9.] How has COVID-19 affected Canada?
\item[10.] Has social distancing had an impact on slowing the spread of COVID-19?
\item[13.] What are the transmission routes of coronavirus?
\item[18.] What are the best masks for preventing infection by Covid-19?
\item[21.] What are the mortality rates overall and in specific populations?
\item[23.] What kinds of complications related to COVID-19 are associated with hypertension?
\item[24.] What kinds of complications related to COVID-19 are associated with diabetes?
\item[26.] What are the initial symptoms of Covid-19?
\item[27.] What is known about those infected with Covid-19 but are asymptomatic?
\item[34.] What are the longer-term complications of those who recover from COVID-19?
\item[43.] How has the COVID-19 pandemic impacted violence in society, including violent crimes?
\item[45.] How has the COVID-19 pandemic impacted mental health?
\item[47.] What are the health outcomes for children who contract COVID-19?
\item[48.] What are the benefits and risks of re-opening schools in the midst of the COVID-19 pandemic?
\end{enumerate}

Notice that the index of the queries match to the original query indices.

\subsection{Doctor-Reviews Queries}
\label{app:doctor_queries_list}

We designed 11 queries with domain experts for the physician review dataset:

\begin{enumerate}
\item How do patients find and choose their doctors?
\item What are patients' experiences with specialist referrals?
\item What breathing problems do patients report and how are they treated?
\item How do doctors manage patients with asthma?
\item What do patients like about their doctors?
\item What do patients dislike about their doctors?
\item What follow-up care or testing do doctors recommend for people with asthma?
\item What do patients like about treatment or management recommendations?
\item What do patients dislike about treatment or management recommendations?
\item What lifestyle challenges do patients with asthma report?
\item What symptoms do patients with asthma report?
\end{enumerate}

\section{Implementation Details: Data Selection Strategies}
\label{app:sampling}

All methods retrieve from the full corpus and select $1,000$ documents per query using fixed random seed ($=42$) for reproducibility. The coding for experiments was assisted by AI tools.

\subsection{Keyword Search}
We use BM25 as our Keyword Search strategy, a lexical retrieval based on BM25Okapi \citep{robertson1994okapi} with parameters $k_1=1.5$ (term frequency saturation) and $b=0.75$ (document length normalization). Documents are preprocessed with Porter stemming and English stopword removal. 

\subsection{SBERT}
SBERT stands for dense semantic retrieval using \texttt{all-mpnet-base-v2}~\citep{reimers-gurevych-2019-sentence} embeddings on queries and documents which retrieves top documents by cosine similarity.

\subsection{Hybrid approaches}
For the following approaches, we retrieve the union of top documents from BM25 and SBERT, re-rank these documents on the hybrid score and select the finally set of documents~(e.g. $1000$ documents for our experiments).

\paragraph{Direct Retrieval} Direct Retrieval uses Simple Sum score that fuses BM25 and SBERT scores by min-max normalizing each independently, then summing for a document d:
\begin{equation*}
\text{score}(d) = \frac{\text{BM25}(d)}{\max(\text{BM25})} + \frac{\text{SBERT}(d)}{\max(\text{SBERT})}
\end{equation*}

\paragraph{Reciprocal Rank Fusion~(RRF)} RRF~\citep{cormack2009reciprocal} is a hybrid approach that combines BM25 and SBERT using rank-based scoring:
\begin{equation*}
\text{RRF}(d) = \frac{1}{\text{rank}_{\text{BM25}}(d) + k} + \frac{1}{\text{rank}_{\text{SBERT}}(d) + k}
\end{equation*}
where $k=60$ is a standard RRF constant. 

% \subsection{Hybrid (Weighted)}

\paragraph{Weighted} Weighted fusion of normalized BM25 and SBERT scores:
\begin{equation*}
\text{score}(d) = w \cdot \frac{\text{BM25}(d)}{\max(\text{BM25})} + (1-w) \cdot \frac{\text{SBERT}(d)}{\max(\text{SBERT})}
\end{equation*}
with $w=0.5$ (equal weighting). 

\subsection{Direct Retrieval + Cross-Encoder}
For this two-stage retrieval, we first use Direct Retrieval to retrieve top-1,000 candidates and then apply cross-encoder model \texttt{cross-encoder/ms-marco-MiniLM-L-6-v2} \citep{nogueira2019passage} to score each (query, document) pair directly to have the final ranking of the 1,000 selected documents.

\subsection{Hybrid + MMR}
We first use direct retrieval to retrieve $5000$ documents and apply Maximal Marginal Relevance score~(MMR)~\citep{carbonell1998use} to iteratively re-rank the documents and select the top $1000$ documents. For a given query embedding $q$, the MMR score for a document embedding $d$ is calculated as follows,

\begin{equation*}
\text{MMR}(d) = \lambda \cdot \text{sim}(d, q) - (1-\lambda) \cdot \max_{d_a \in A} \text{sim}(d, d_a)
\end{equation*}
where $\lambda\in [0,1]$ is the parameter controlling diversity strength and $A$ is the set of already-selected documents. We follow prior works to set $\lambda=0.3$ balancing relevance and diversity, and similarity metric is cosine similarity.

\subsection{Query Expansion (KeyBERT)}

\paragraph{Stage 1 (Keyword Extraction):} We extract 5 keywords from the query using KeyBERT \citep{grootendorst2020keybert} with \texttt{all-mpnet-base-v2} embeddings and diversity parameter 0.7 for MMR-based keyword selection.

\paragraph{Stage 2 (Expanded Retrieval):} We apply Direct Retrieval for the original query as well as for each of the 5 keywords, establishing 6 retrievals in total. The documents are selected based on a weighted RRF score from the six retrievals:
\begin{equation*}
\text{RRF}(d) = \frac{w_{orig}}{k + \text{rank}_{\text{orig}}(d)} + \sum_{i=1}^{5} \frac{w_i}{k + \text{rank}_{i}(d)}
\end{equation*}
where $k = 60$ is a standard RRF constant and weights are set to $w_{orig}=0.5, w_i = 0.1 \forall i$ such that queries using expanded keywords share weights equally. 

\subsection{Retrieval Random}
This selection initially retrieves 5,000 documents via Direct Retrieval, then uniformly samples 1,000 documents (Seed for reproducibility: 42). 
% Removes ranking bias while maintaining relevance filtering, testing whether retrieval filtering alone affects topic modeling.

\begin{figure*}[htb]
    \centering
    \includegraphics[width=0.49\textwidth]{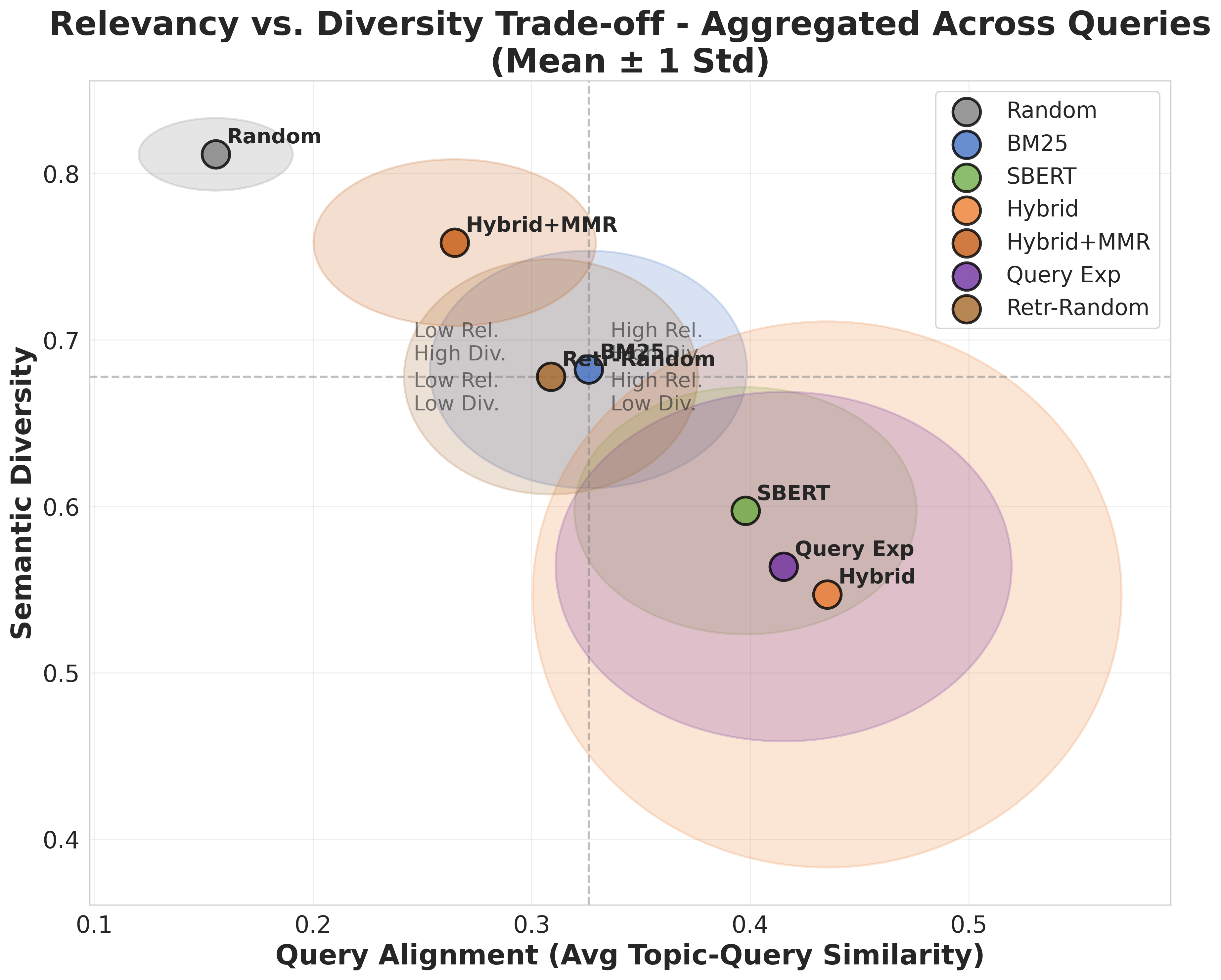}
    \includegraphics[width=0.49\textwidth]{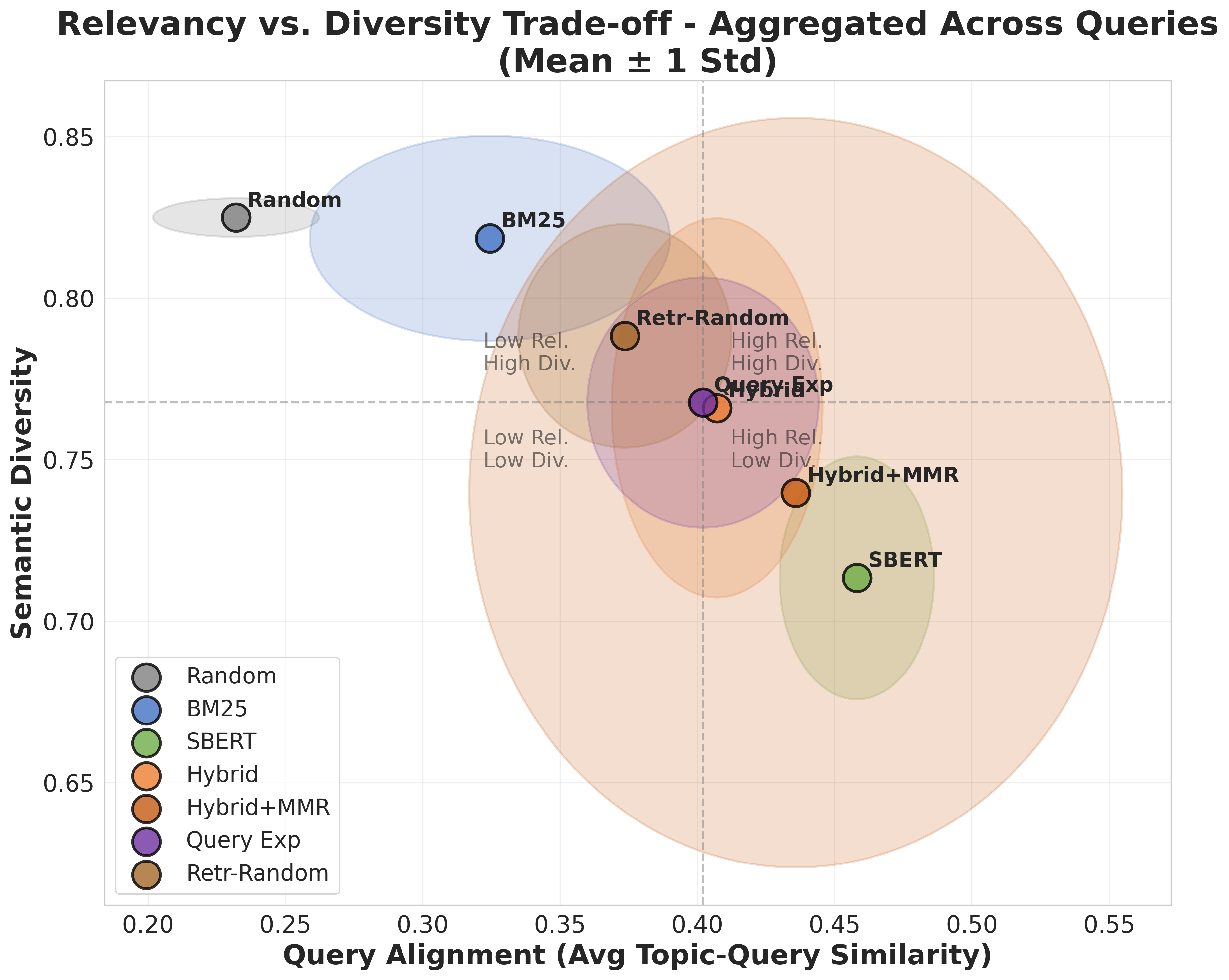}
    \includegraphics[width=0.49\textwidth]{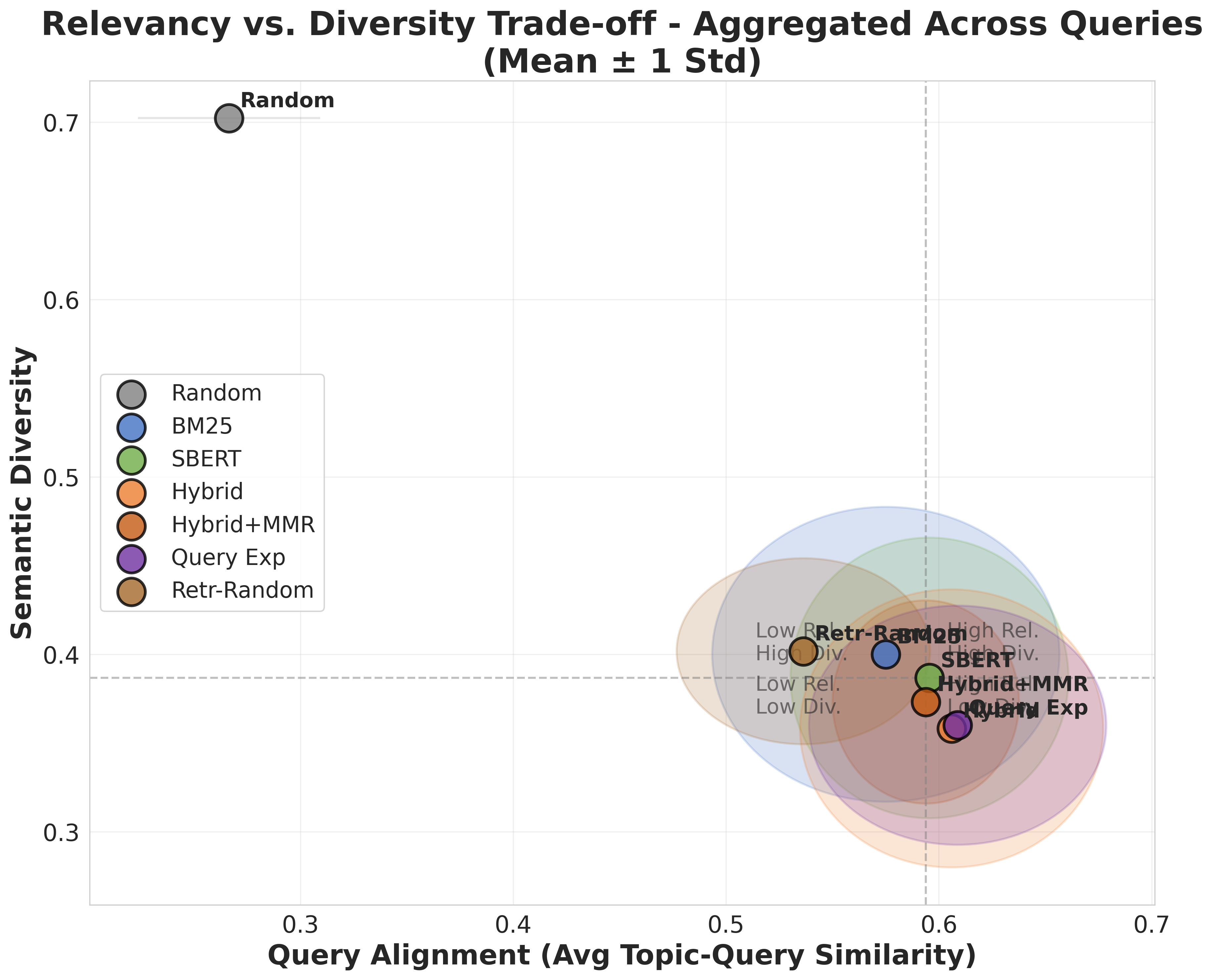}
    \includegraphics[width=0.49\textwidth]{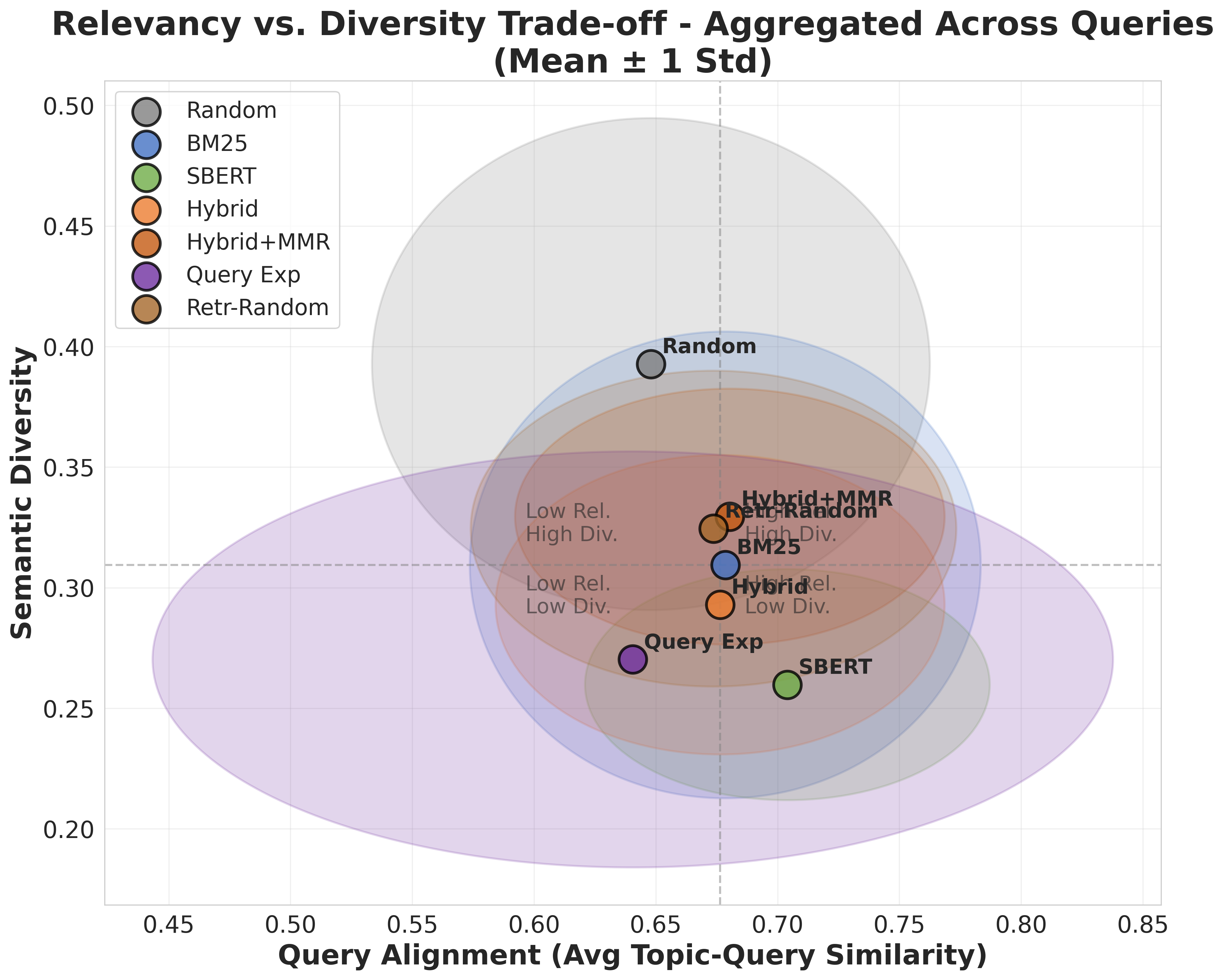}
    \caption{\textbf{Relevance-diversity trade-off on TREC-COVID (15 queries).} Topic-query similarity versus overall semantic diversity for TopicGPT (top-left), BERTopic (top-right), LDA (bottom-left), and HiCode (bottom-right). All models show negative correlation: higher relevance → lower diversity. Random Uniform achieves highest diversity but lowest relevance. Semantic methods (SBERT, Direct Retrieval, Query Expansion) cluster in high-relevance, moderate-diversity region. TopicGPT and BERTopic show most pronounced trade-offs with large diversity reduction (TopicGPT: 51\%, BERTopic: 49\%). LDA shows steeper diversity reduction (49\%) across narrow alignment range. HiCode shows compressed relevance range due to query-aware generation.}
    \label{fig:diversity_trec_all}
\end{figure*}

\section{Implementation Details: Topic Modeling}
\label{app:topic_models}

We perform the same pre-processing for LDA and BERTopic to build vocabulary using the setting: \texttt{min\_df=2}, \texttt{max\_df=0.95}, unigrams and bigrams.

\paragraph{Topic Count Matching} For each data selection strategy, given a query, we first run BERTopic to determine discovered topic count (excluding outliers), then configure LDA to use the same number for the number of topics. If BERTopic results are unavailable, we set LDA to cluster 20 topics. 

\paragraph{Topic representation} For BERTopic and LDA,  we concatenate top-10 words by c-TF-IDF score (BERTopic) or topic-word probability (LDA), encode concatenated string to embedding. For TopicGPT and HiCode, we directly encode natural language topic description/label to embeddings.

\subsection{LDA}
We implement LDA model with \texttt{gensim} package's \texttt{LdaMulticore}~\citep{vrehuuvrek2010software} with the following hyperparameters: $\alpha=\text{``symmetric''}$ (uniform document-topic prior), $\eta=0.01$ (sparse topic-word prior for focused topics), 15 passes through corpus, up to 100 iterations per pass, and 20 worker threads for parallelization.

\subsection{BERTopic}
BERTopic consists of three major modules: embedding, dimention reduction, and clustering. For embedding, we keep using \texttt{all-mpnet-base-v2}. Dimension reductions employs UMAP with the following parameters: \texttt{n\_neighbors=15}, \texttt{n\_components=5}, \texttt{min\_dist=0.0}, and \texttt{metric=cosine}. Finally, we use HDBSCAN in clustering and set \texttt{min\_cluster\_size=5}, \texttt{metric=euclidean}, and \texttt{cluster\_selection\_method=eom} (excess of mass). Documents not assigned to any cluster marked as outliers where their topic ID is labeled as $-1$.

\subsection{TopicGPT}
We implement TopicGPT as follows. \textbf{Stage 1 (Topic Generation):} We randomly sample 500 documents as the input for topic generation using \texttt{gpt-4o-mini} with temperature set to zero. \textbf{Stage 2 (Topic Assignment):} Assign all 1,000 documents to generated topics using \texttt{gpt-4o-mini} with \texttt{temperature=0} using structured prompt including complete topic hierarchy and descriptions. 

\textbf{Cost:} Approximately \$0.50--\$2.00 per query-method pair using gpt-4o-mini (\$0.15/1M input tokens, \$0.60/1M output tokens).

\subsection{HiCode}
We apply HiCode's hierarchical prompting modules using \texttt{gpt-4o-mini} with temperature $=0$. HiCode has a query-aware topic generation that explicitly incorporates query texts into prompts at each hierarchical level and instructs the LLM to produce topics relevant to the query. It generates 2--4 levels of topic hierarchy. We iteratively assign the final topics back to the documents.

\textbf{Query-Aware Architecture Note:} HiCode's explicit query incorporation explains the convergence in alignment values across selection methods (Tab.~\ref{tab:alignment_all_models}). The model biases topic discovery toward query concepts regardless of sample composition. This architectural choice makes HiCode highly effective for query-focused topic discovery but masks selection method differences in comparative evaluation.

\begin{figure*}[htb]
    \centering
    \includegraphics[width=0.49\textwidth]{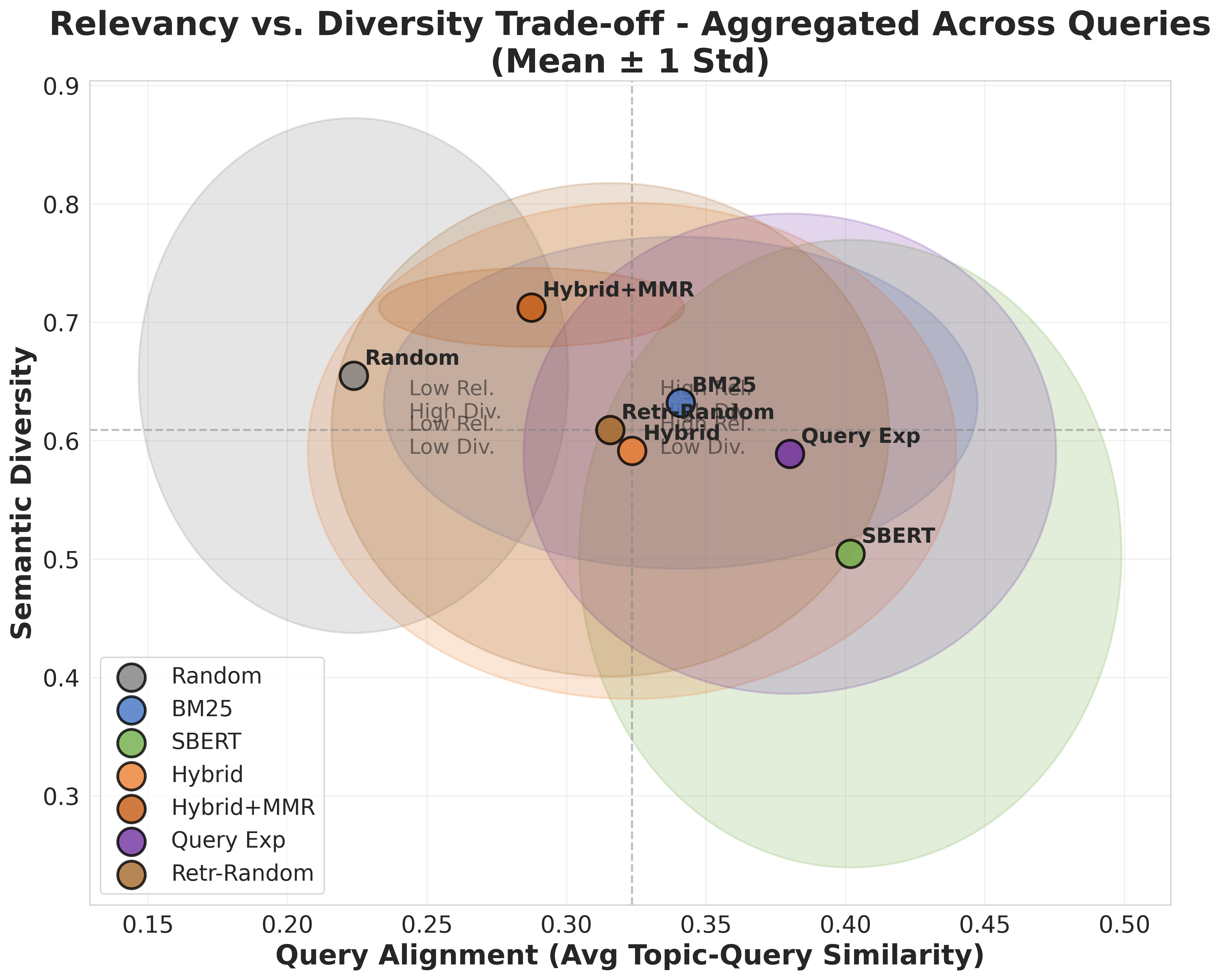}
    \includegraphics[width=0.49\textwidth]{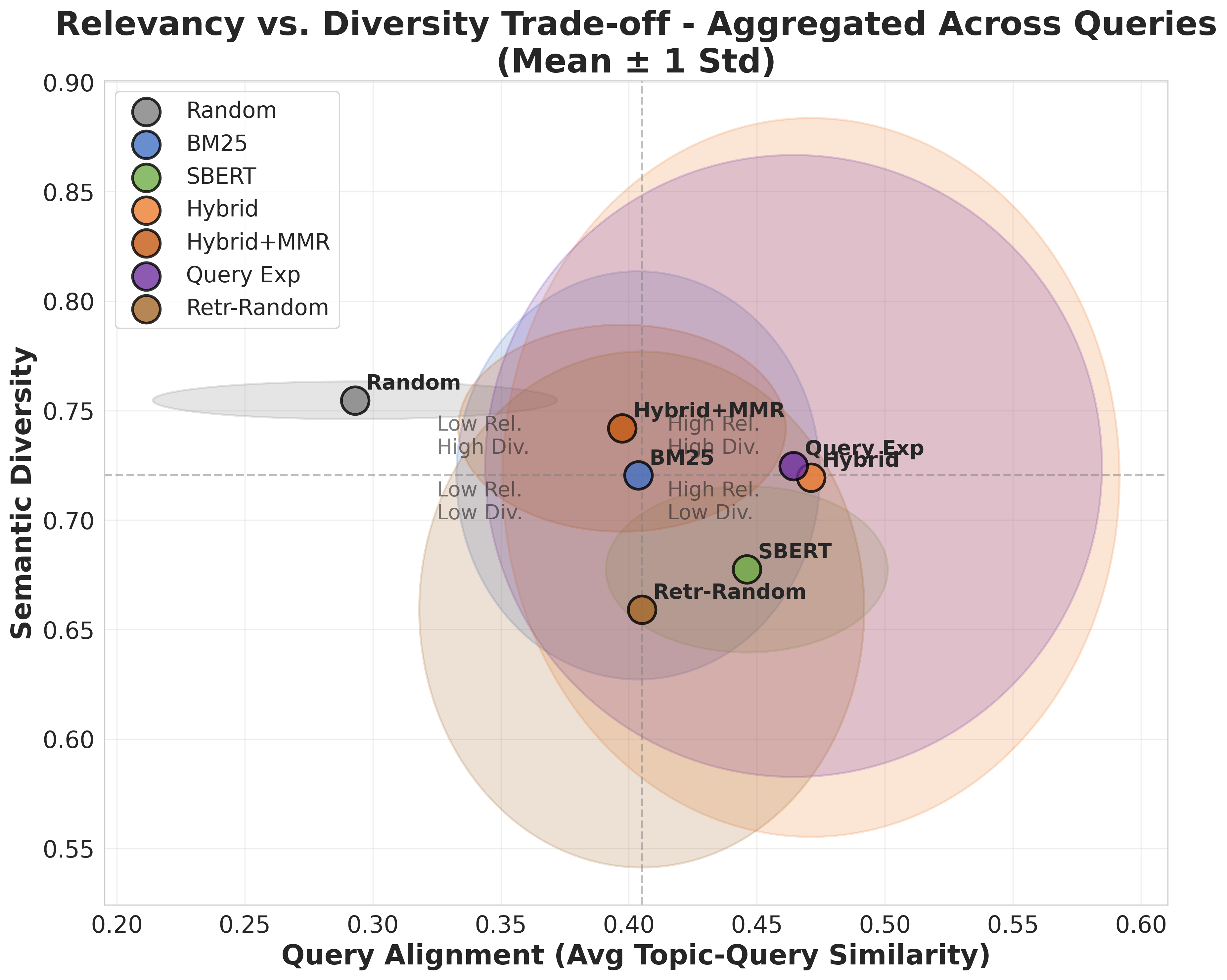}
    \includegraphics[width=0.49\textwidth]{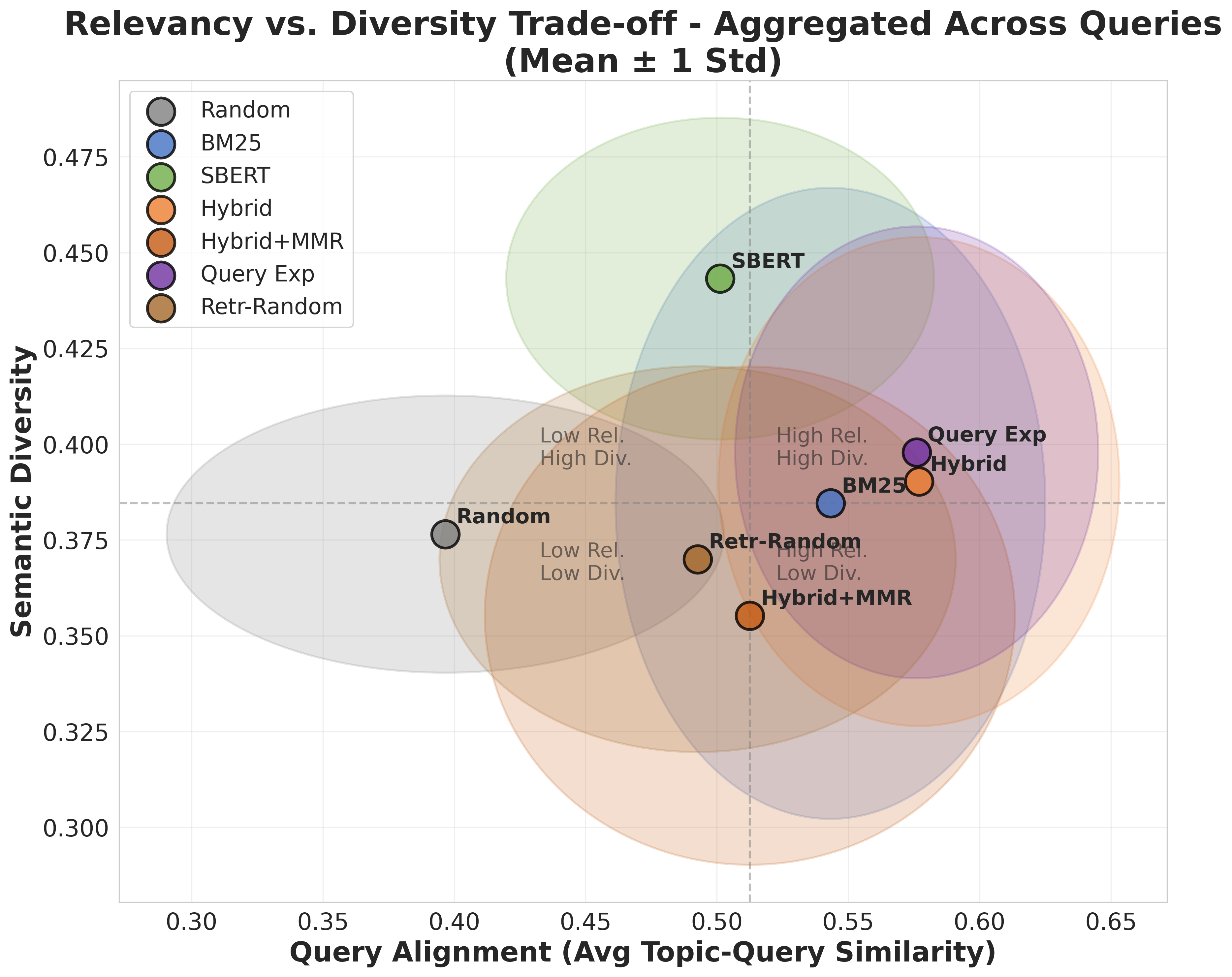}
    \includegraphics[width=0.49\textwidth]{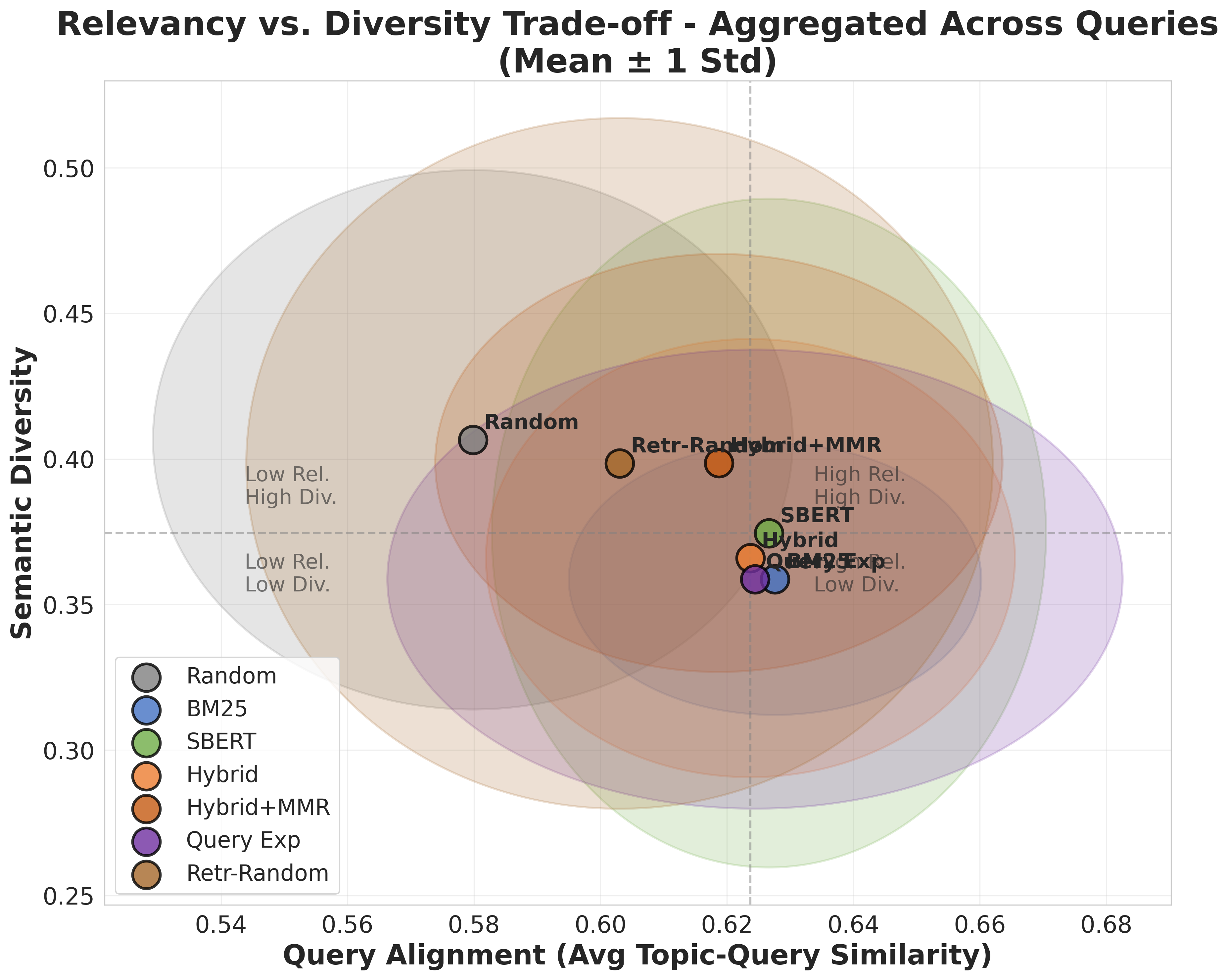}
    \caption{\textbf{Relevance-diversity trade-off on Doctor-Reviews (11 queries).} TopicGPT (top-left), BERTopic (top-right), LDA (bottom-left), and HiCode (bottom-right) all show negative correlation but with attenuated differentiation compared to TREC-COVID. Higher variance and overlapping error bars reflect weaker query quality (App.~\ref{app:doctor_queries}). HiCode maintains compressed ranges consistent with query-aware generation.}
    \label{fig:diversity_doctor_all}
\end{figure*}

\begin{figure*}[t]
    \centering
    \includegraphics[width=0.49\textwidth]{figures/topicgpt_trec_relevant_diversity.png}
    \includegraphics[width=0.49\textwidth]{figures/bertopic_trec_relevant_diversity.png}
    \includegraphics[width=0.49\textwidth]{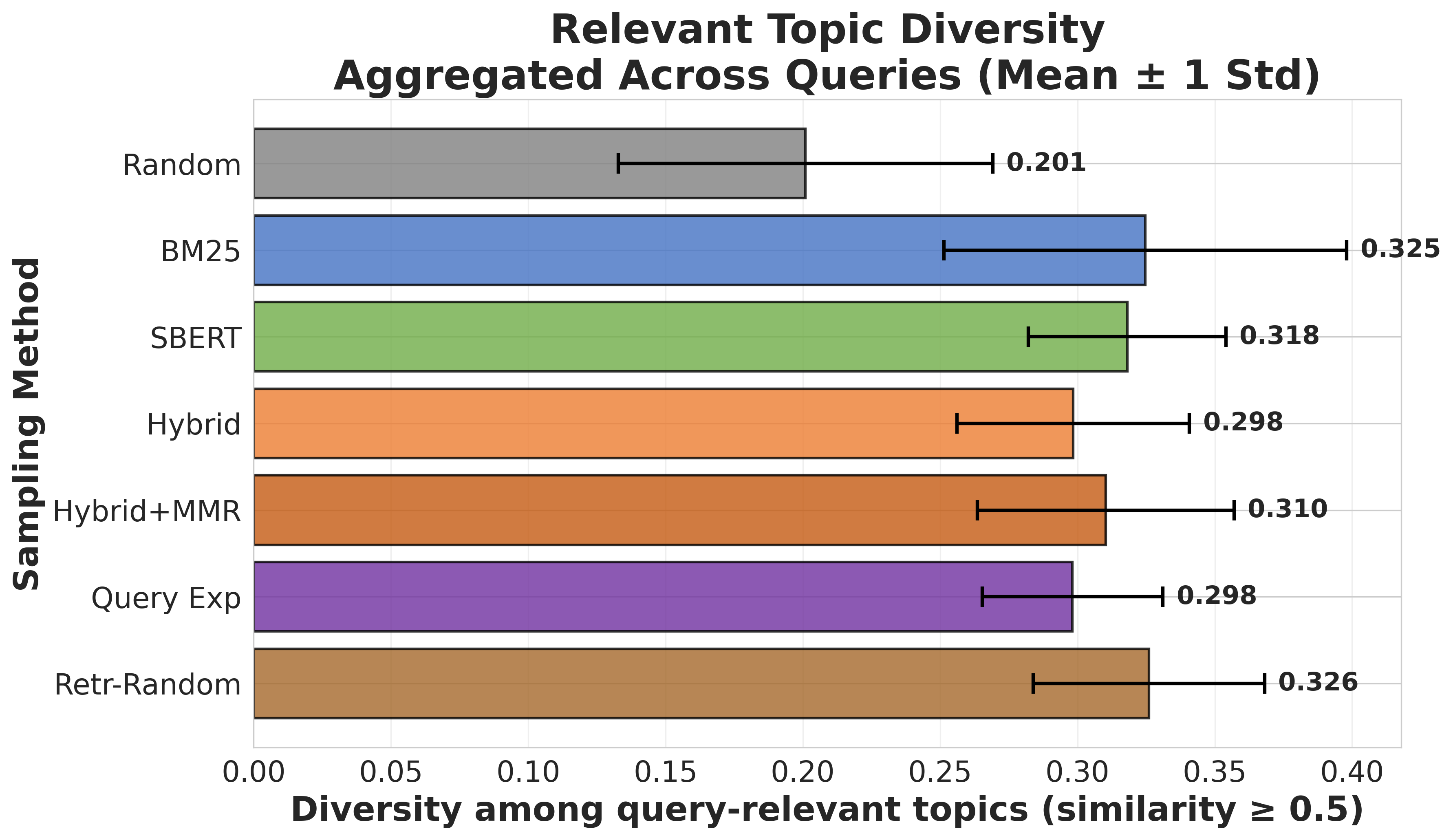}
    \includegraphics[width=0.49\textwidth]{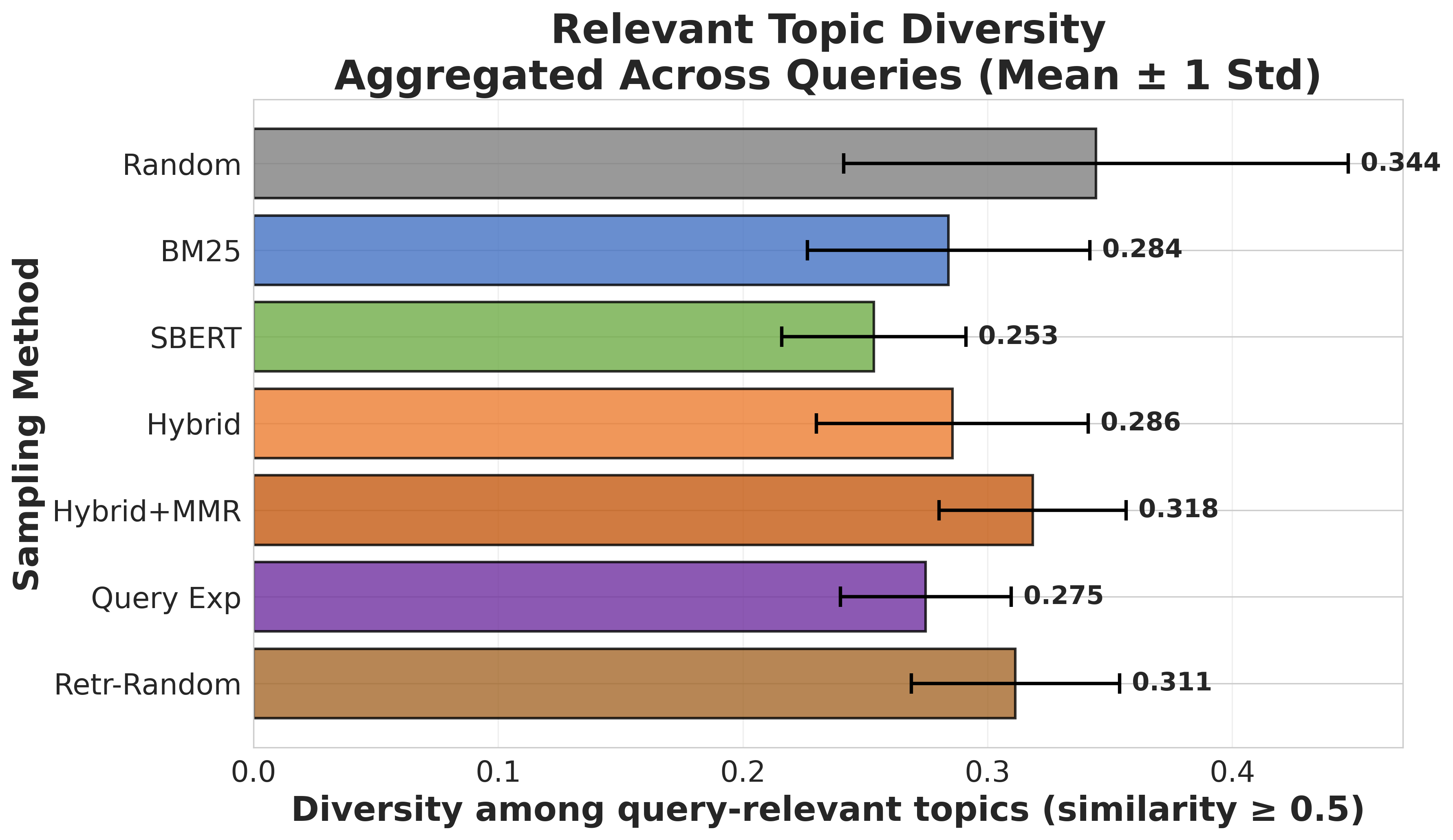}
    \caption{\textbf{Diversity among query-relevant topics only on TREC-COVID (15 queries).} TopicGPT (top-left), BERTopic (top-right), LDA (bottom-left), and HiCode (bottom-right). Error bars show $\pm$1 standard deviation. All models show overlapping values across methods: TopicGPT 0.416--0.476, BERTopic 0.386--0.484, LDA 0.447--0.481, HiCode 0.523--0.558. This demonstrates that while overall diversity decreases with alignment (Fig.~\ref{fig:diversity_trec_all}), diversity is preserved among topics semantically related to the query. High overall diversity in Random Uniform primarily reflects off-topic content rather than richer query-relevant coverage.}
    \label{fig:relevant_div_trec_all}
\end{figure*}

\begin{figure*}[htbp]
    \centering
    \includegraphics[width=0.49\textwidth]{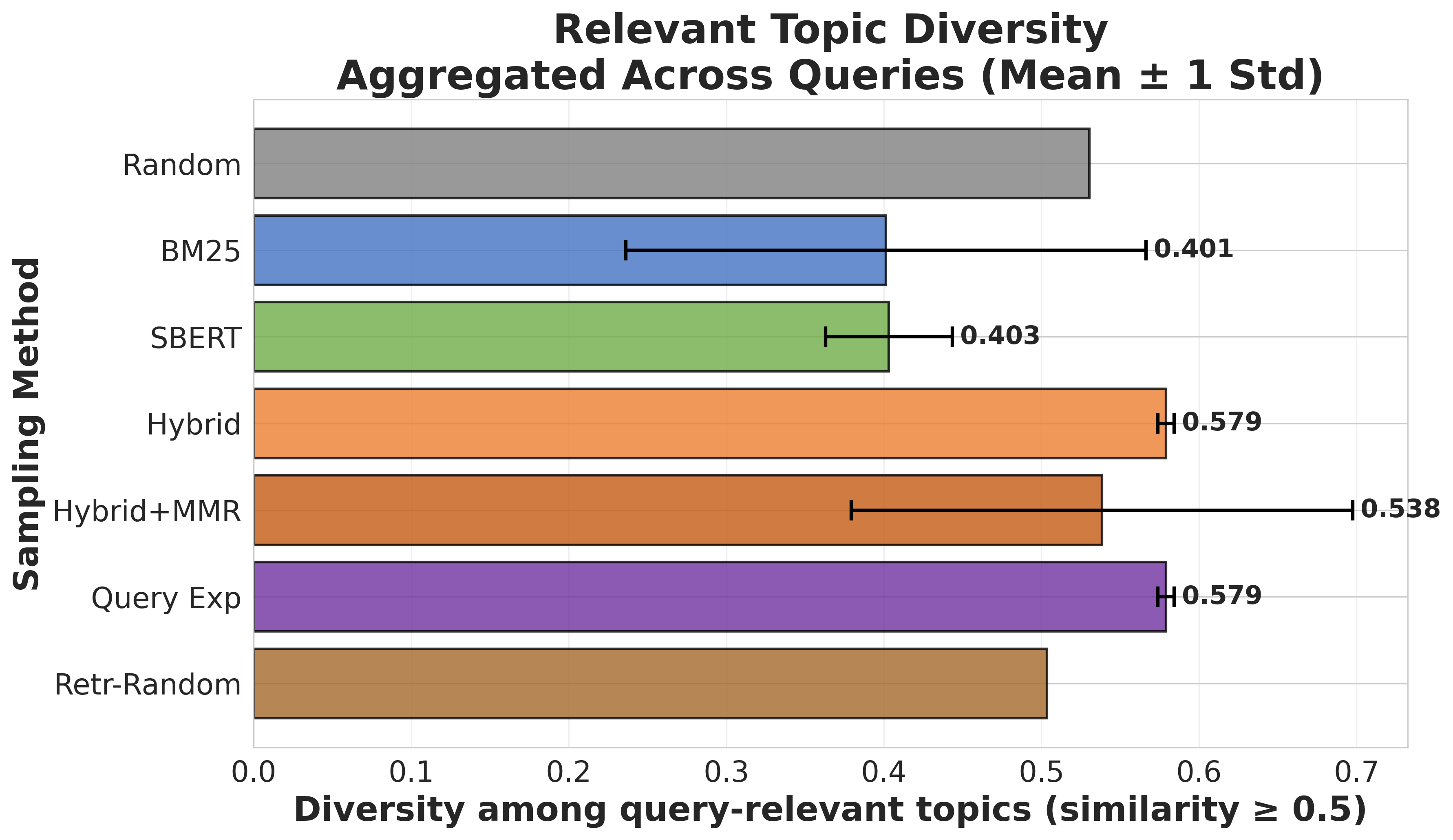}
    \includegraphics[width=0.49\textwidth]{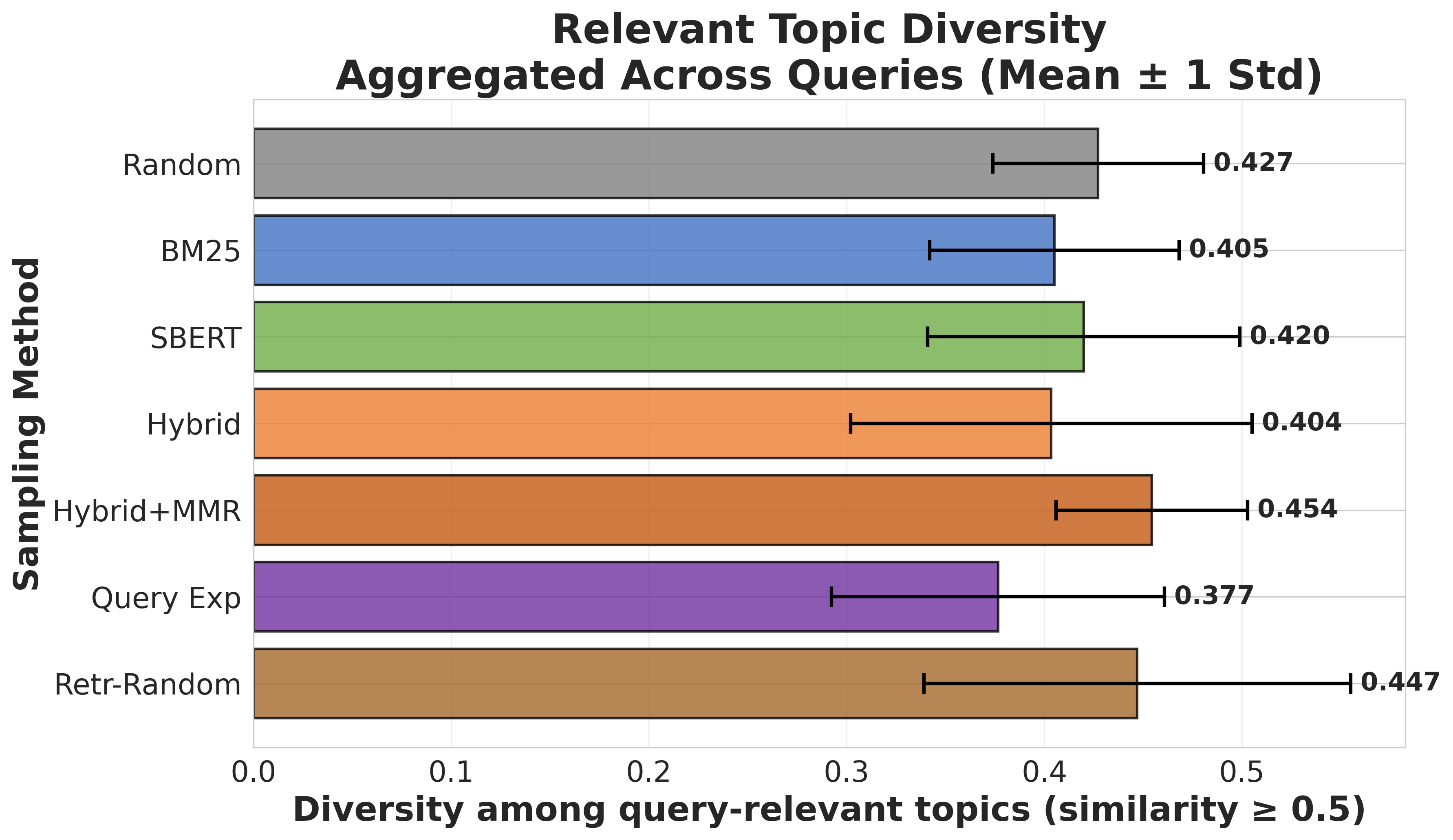}
    \includegraphics[width=0.49\textwidth]{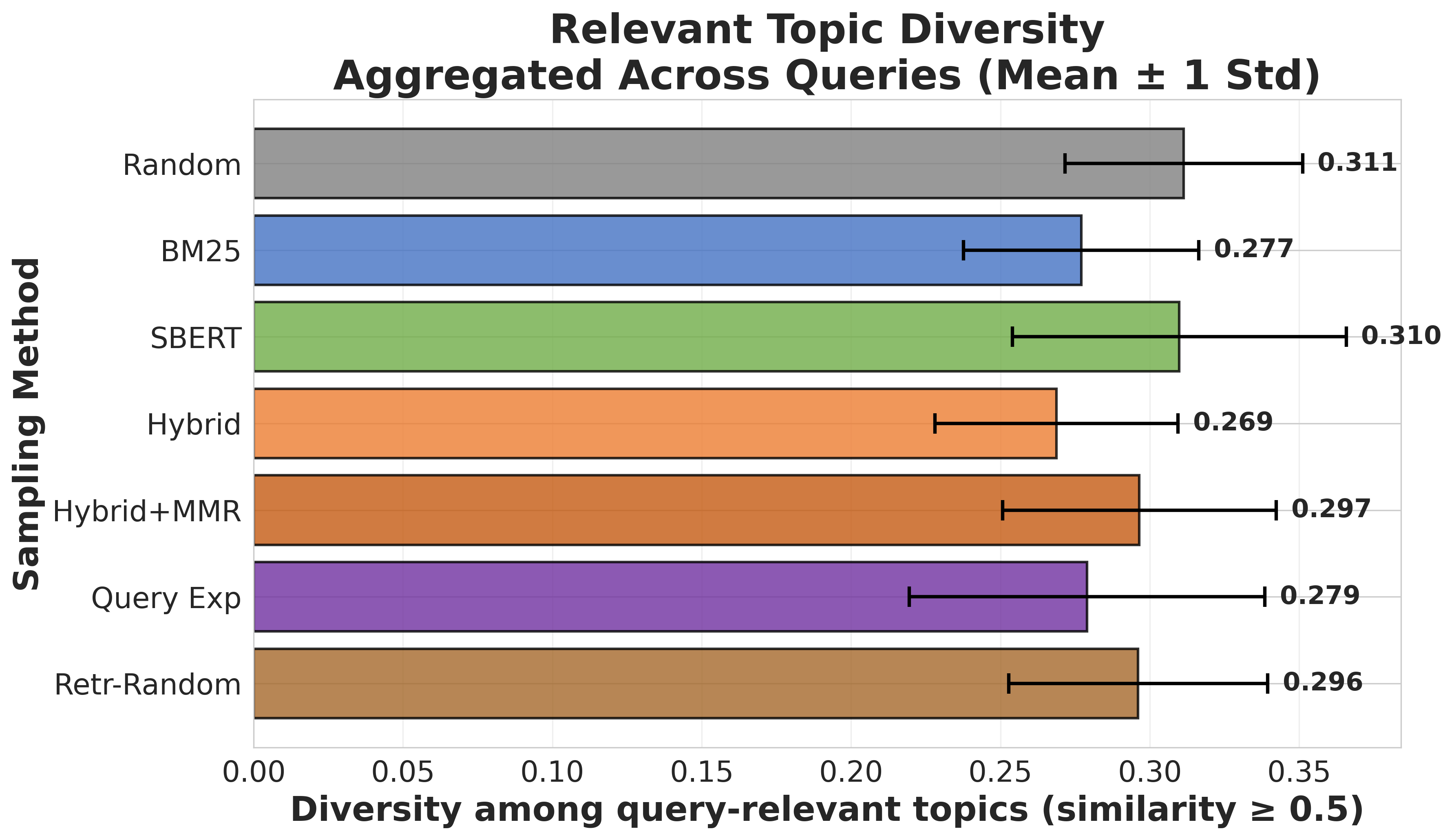}
    \includegraphics[width=0.49\textwidth]{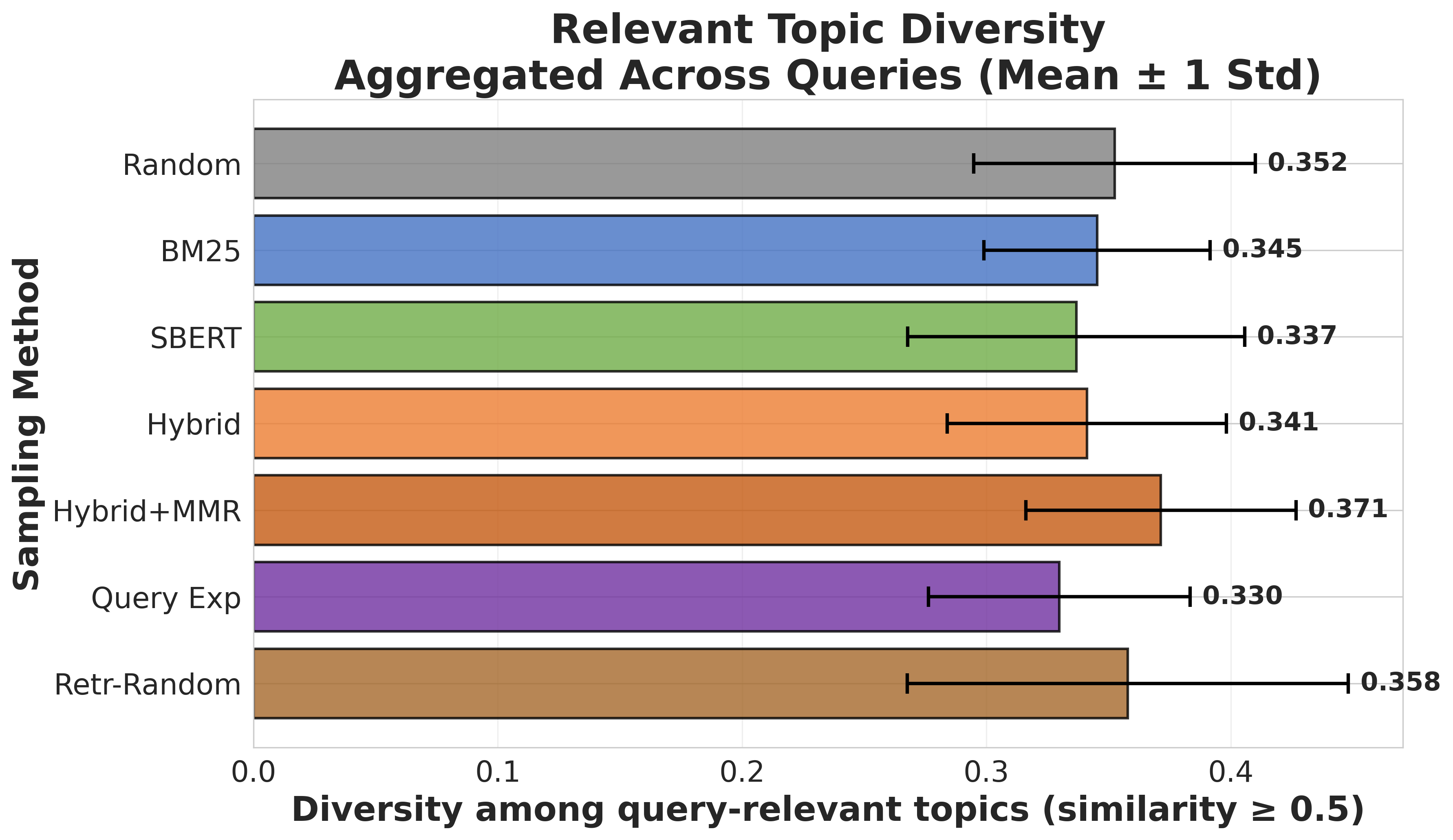}
    \caption{\textbf{Diversity among query-relevant topics only on Doctor-Reviews (11 queries).} All four models, TopicGPT (top-left), BERTopic (top-right), LDA (bottom-left), and HiCode (bottom-right), show overlapping ranges with minimal differentiation across selection methods. Pattern consistent with TREC-COVID (Fig.~\ref{fig:relevant_div_trec_all}): retrieval preserves diversity among query-relevant topics while filtering off-topic content. Slightly higher variance reflects weaker query quality (App.~\ref{app:doctor_queries}).}
    \label{fig:relevant_div_doctor_all}
\end{figure*}
\section{Topic Diversity: Supplementary Figures}
\label{app:diversity_all}
\subsection{Overall Diversity}
\paragraph{TREC-COVID}The relevance-diversity trade-off appears consistently across all four topic models on TREC-COVID (Fig.~\ref{fig:diversity_trec_all}), though magnitude varies. TopicGPT and BERTopic exhibit the most pronounced trade-offs, with Random Uniform achieving 51\% and 49\% higher diversity than semantic methods (SBERT, Direct Retrieval, Query Expansion) respectively. LDA shows a comparably steep reduction (49\%) across a narrower relevance range (0.266--0.609), reflecting its bag-of-words probabilistic framework where lexical overlap more directly impacts both relevance scoring and topic structure. HiCode displays compressed relevance ranges (0.641--0.704) with minimal diversity variation, as query-aware generation biases all discovered topics toward query concepts regardless of sample composition. 
Across all selection methods, semantic methods cluster in the high-relevance, moderate-diversity quadrant, while Random Uniform consistently occupies the low-relevance, high-diversity corner, though this diversity includes substantial off-topic content as established in Sec.~\ref{sec:query_alignment}.

\paragraph{Doctor-Reviews} Similarly, in Fig.~\ref{fig:diversity_doctor_all},  Doctor-Reviews exhibits weaker relevance-diversity differentiation compared to TREC-COVID, with all four models showing attenuated slopes and higher variance. The negative correlation persists but the clustering of selection methods is less distinct, with substantial error bar overlap across semantic methods. This pattern aligns with the query quality analysis (App.~\ref{app:doctor_queries}): when queries are vague or compound (Query 3: mean relevance 0.450), selection methods provide minimal differentiation. HiCode maintains compressed ranges (0.603--0.628) consistent with TREC-COVID, though absolute relevance values are lower. The higher variance reflects heterogeneous query quality: strong queries (6, 2, 7) show TREC-COVID-like differentiation, while weak queries (3, 10, 11) approach random performance.

\subsection{Relevant Topic Diversity}
\label{app:relevant_div_all}

\paragraph{TREC-COVID} When diversity is computed only among query-relevant topics (similarity $\geq 0.5$), differentiation between selection methods disappears across all models (Fig.~\ref{fig:relevant_div_trec_all}). TopicGPT shows overlapping ranges of 0.416--0.476, BERTopic 0.386--0.484, LDA 0.447--0.481, and HiCode 0.523--0.558, with substantial error bar overlap indicating no meaningful separation. This contrasts sharply with overall diversity (Fig.~\ref{fig:diversity_trec_all}), where Random Uniform achieves 35--51\% higher diversity than semantic methods. The implication: diversity ``lost'' through retrieval primarily affects off-topic or tangentially related topics, not query-relevant topics. Random Uniform's high overall diversity reflects noise rather than richer query-relevant coverage. Semantic methods preserve diversity where it matters- among topics semantically related to the information need, while filtering irrelevant content. Random Uniform's lower relevant diversity in TopicGPT reflects scarcity: with few relevant topics discovered, diversity calculations become unstable.

\paragraph{Doctor-Reviews} Doctor-Reviews shows consistent patterns with TREC-COVID: relevant topic diversity exhibits minimal differentiation across selection methods for all four models (Fig.~\ref{fig:relevant_div_doctor_all}), with TopicGPT ranges 0.387--0.442, BERTopic 0.356--0.421, LDA 0.398--0.445, and HiCode 0.421--0.468. Error bars overlap substantially, indicating preserved diversity among query-relevant topics regardless of selection strategy. The slightly higher variance compared to TREC-COVID reflects weaker query quality in some queries (App.~\ref{app:doctor_queries}), where even semantic methods struggle to differentiate from random sampling (Query 3: 0.450 mean alignment). Nonetheless, the core finding holds: retrieval bias does not reduce diversity among topics that matter for the query.

\section{Pairwise Coverage Analysis: Supplementary Figures and Tables}
\label{app:coverage_all}

\begin{figure*}[htbp]
    \centering
    \includegraphics[width=0.49\textwidth]{figures/topicgpt_trec_coverage_heatmap.png}
    \includegraphics[width=0.49\textwidth]{figures/bertopic_trec_coverage_heatmap.png}
    \includegraphics[width=0.49\textwidth]{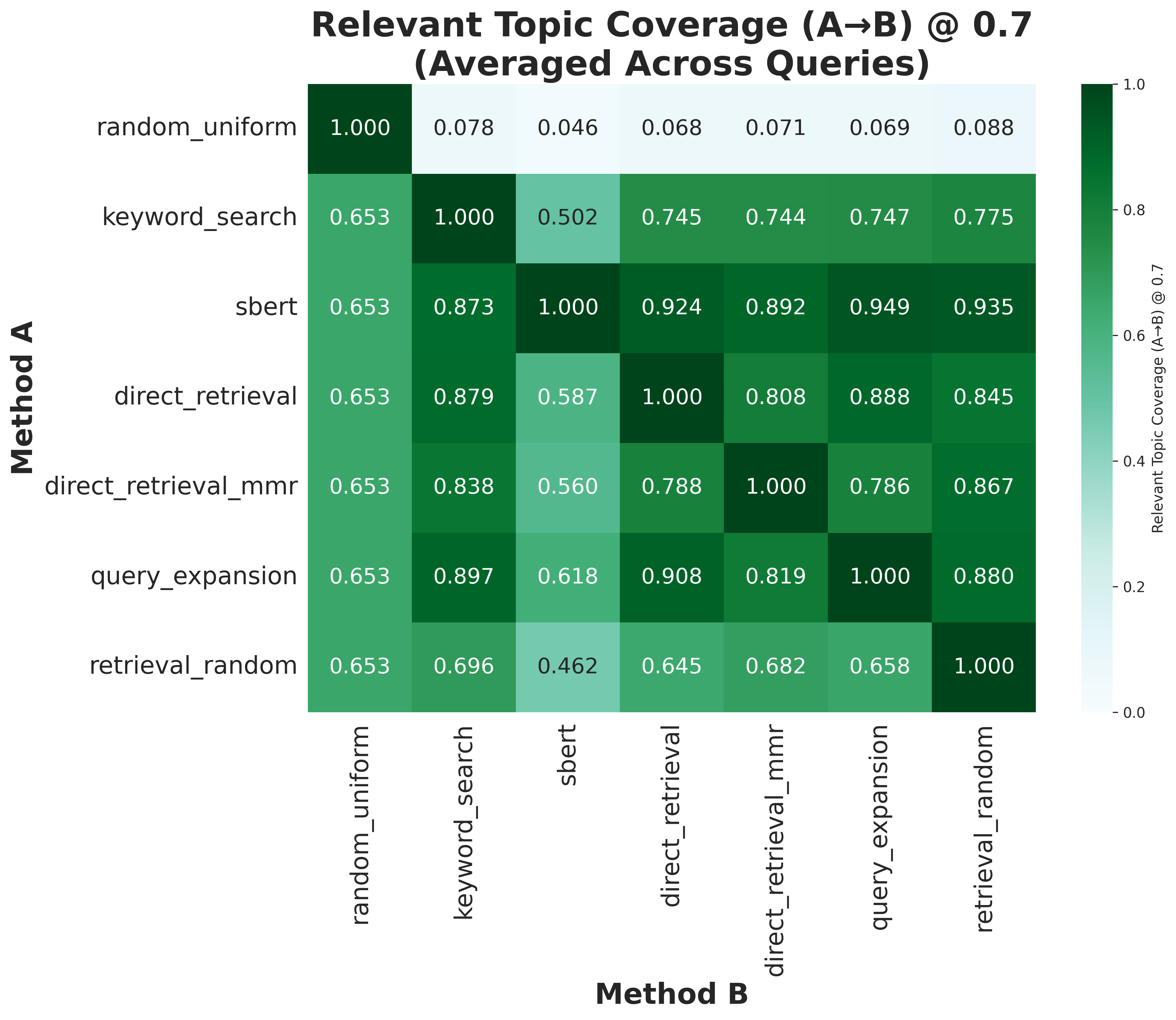}
    \includegraphics[width=0.49\textwidth]{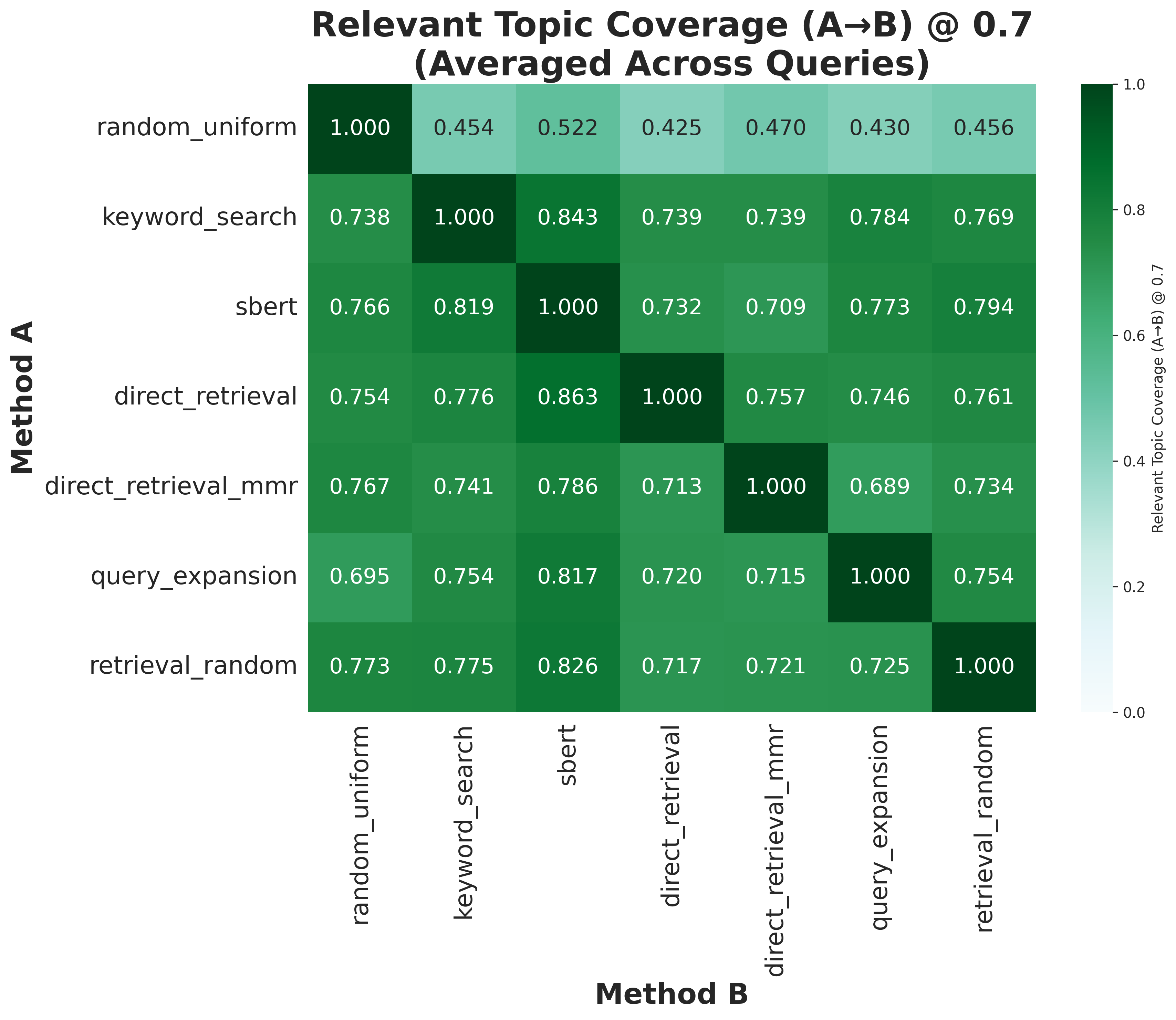}
    \caption{\textbf{Pairwise relevant topic coverage at threshold 0.7 on TREC-COVID (15 queries).} The plots top-to- bottom left-to-right are TopicGPT, BERTopic, LDA, HiCode. Cell $(i,j)$ shows fraction of method $j$'s relevant topics covered by method $i$'s relevant topics. Semantic methods (SBERT, Direct Retrieval, Query Expansion) show dark rows (0.52--0.68 coverage), indicating convergence on similar query-relevant topics. Random Uniform shows very light rows (0.04--0.09), discovering predominantly off-topic content. LDA (bottom-left) shows substantially higher coverage across all methods (0.65--0.83 among semantic methods), reflecting structural convergence discussed in Sec.~\ref{sec:discussion}. HiCode (bottom-right) shows high symmetric coverage (0.70--0.85) due to query-aware generation.}
    \label{fig:coverage_trec_all}
\end{figure*}

\paragraph{TREC-COVID} Fig.~\ref{fig:coverage_trec_all} shows pairwise relevant topic coverage at similarity threshold 0.7 for all topic models. It reveals convergence patterns across all selection strategies. Semantic methods (SBERT, Direct Retrieval, Query Expansion) show dark rows (TopicGPT: 0.52--0.64, BERTopic: 0.54--0.68 coverage), indicating they discover similar sets of query-relevant topics despite distinct sample compositions. Random Uniform shows very light rows across all methods (0.04--0.09), confirming it discovers predominantly off-topic content. LDA exhibits notably higher coverage across all methods (0.65--0.83 among semantic methods), reflecting structural convergence in its probabilistic topic-word distributions, when applied to relevance-filtered samples, LDA's Dirichlet priors guide it toward similar latent structures regardless of selection nuances. HiCode shows high symmetric coverage (0.70--0.85) driven by query-aware generation: explicit query incorporation in hierarchical prompts biases all methods toward the same query-aligned concepts.

\begin{figure*}[t]
    \centering
    \includegraphics[width=0.49\textwidth]{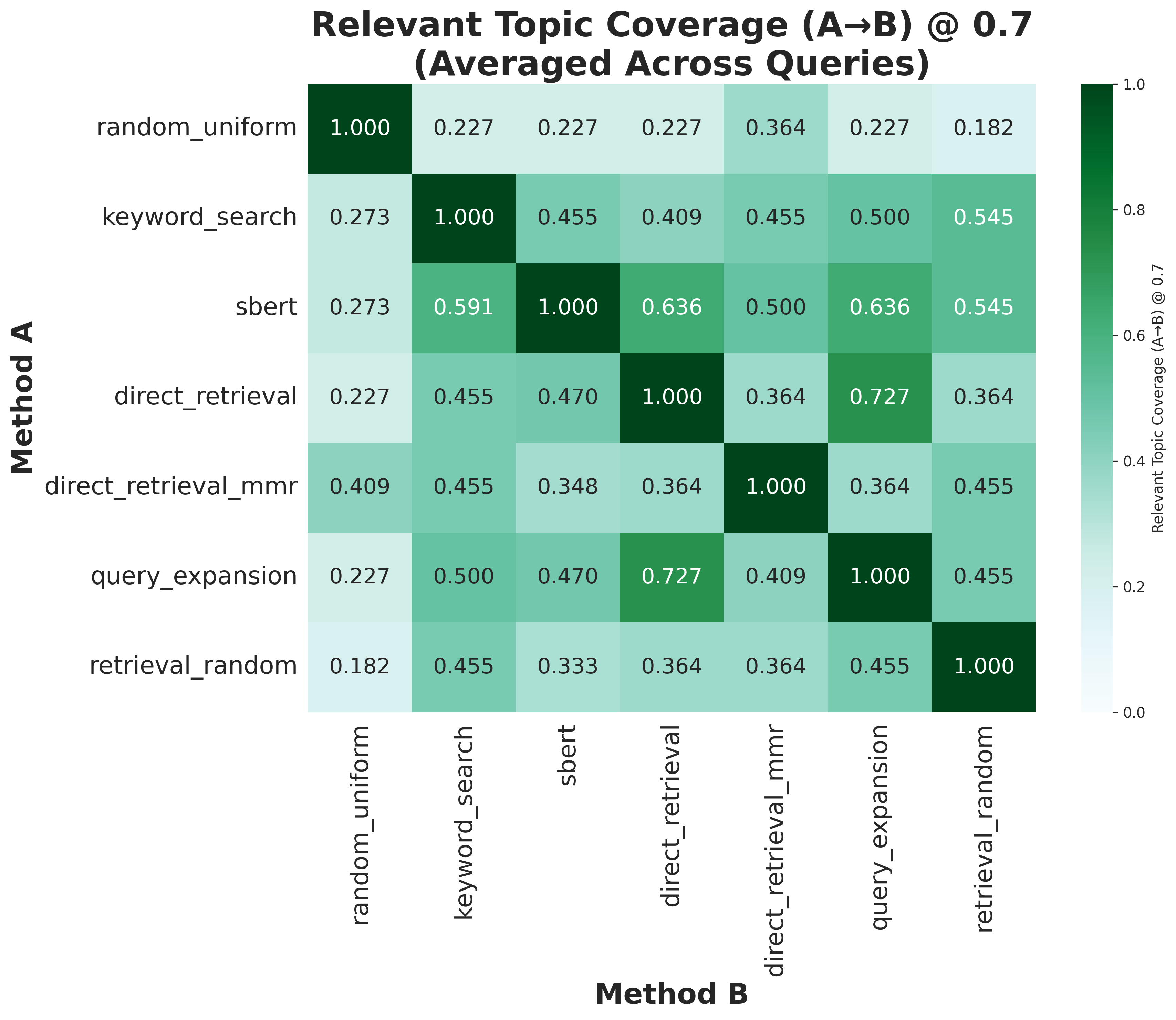}
    \includegraphics[width=0.49\textwidth]{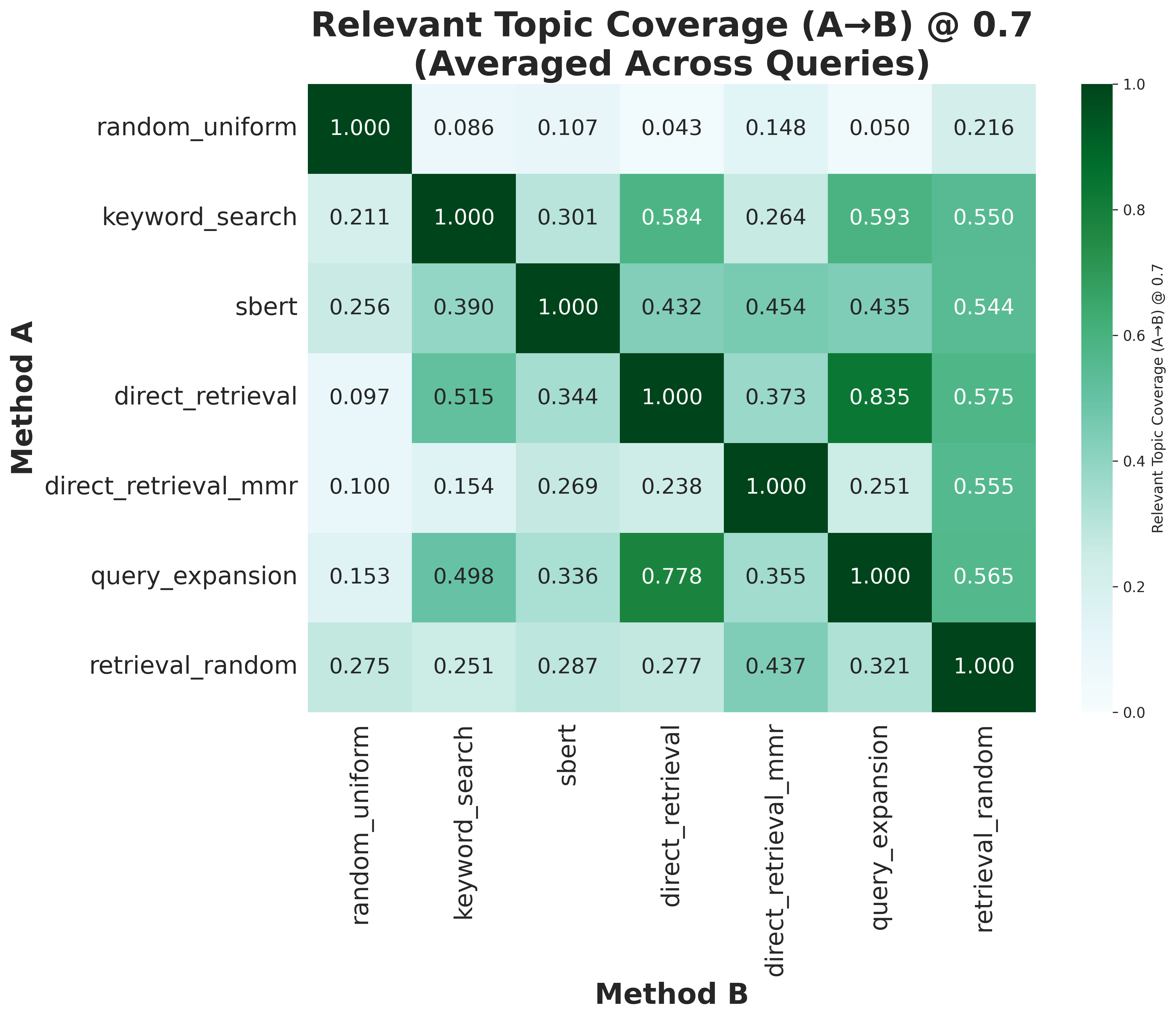}
    \includegraphics[width=0.49\textwidth]{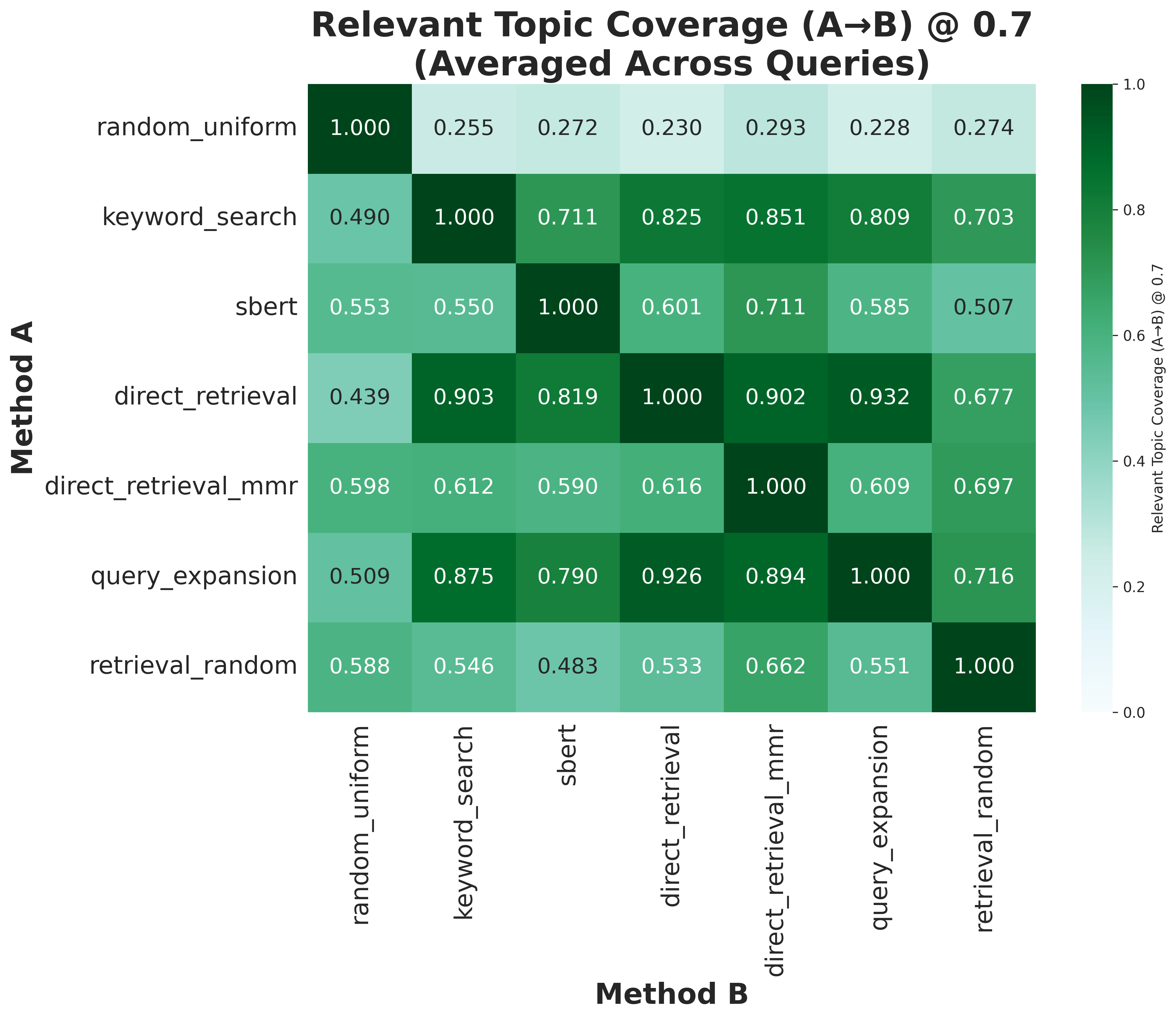}
    \includegraphics[width=0.49\textwidth]{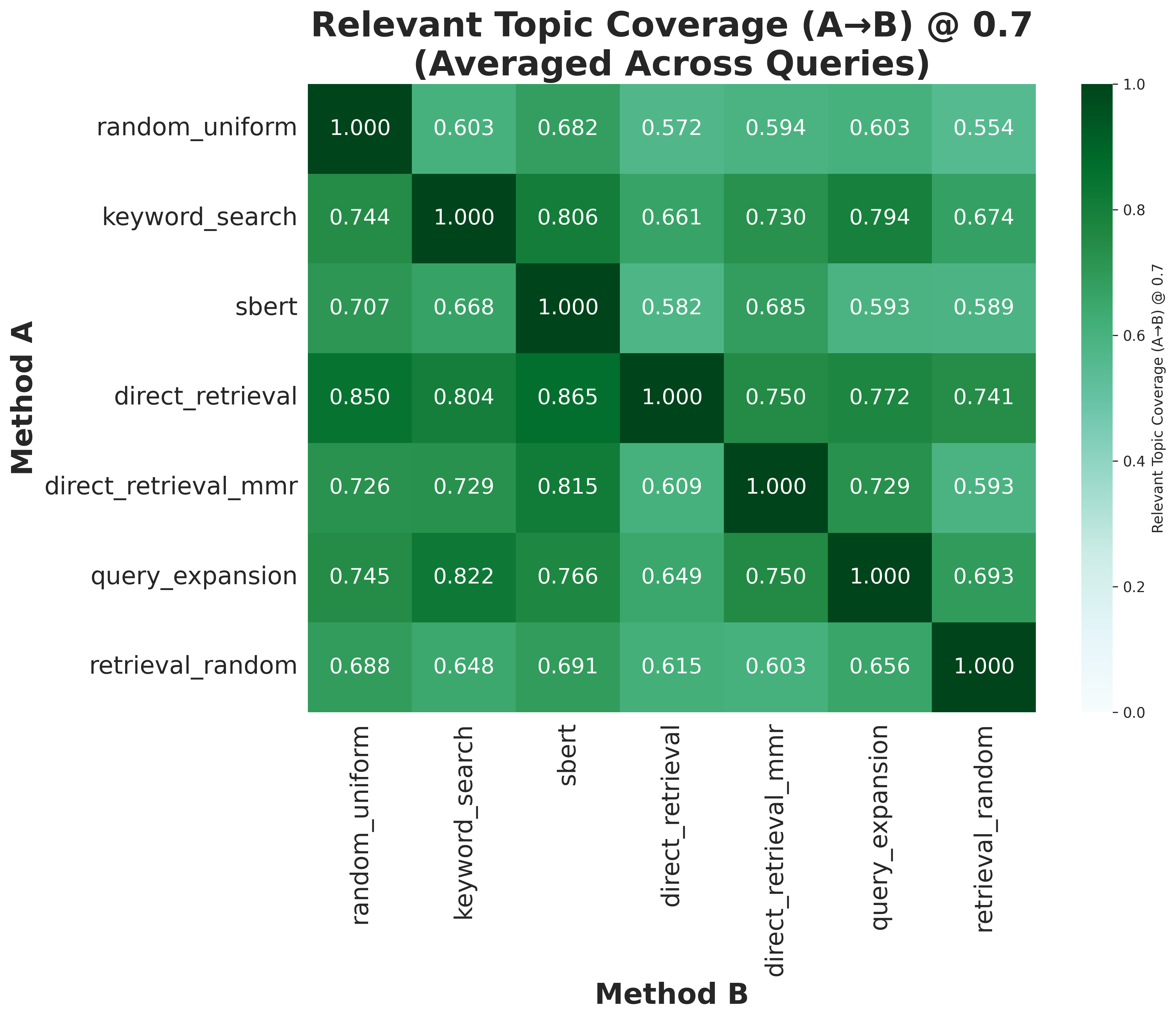}
    \caption{\textbf{Pairwise relevant topic coverage at threshold 0.7 on Doctor-Reviews (11 queries).} Cell $(i,j)$ shows fraction of method $j$'s relevant topics covered by method $i$. All four models, TopicGPT (top-left), BERTopic (top-right), LDA (bottom-left), and HiCode (bottom-right), show semantic methods with darker rows than Random Uniform, indicating convergence on similar relevant topics. Patterns consistent with TREC-COVID (Fig.~\ref{fig:coverage_trec_all}) but with higher variance reflecting query quality heterogeneity (App.~\ref{app:doctor_queries}).}
    \label{fig:coverage_doctor_all}
\end{figure*}

\paragraph{Doctor-Reviews} Doctor-Reviews coverage patterns mirror TREC-COVID but with higher variance (Fig.~\ref{fig:coverage_doctor_all}). Semantic methods maintain darker rows than Random Uniform across all four topic models, though differentiation is less pronounced. TopicGPT and BERTopic show 0.43--0.61 coverage among semantic methods (vs. 0.52--0.68 on TREC-COVID), while LDA maintains high coverage (0.58--0.76) consistent with cross-dataset structural convergence. HiCode shows 0.62--0.78 coverage, slightly lower than TREC-COVID but maintaining the query-aware convergence pattern. Higher variance reflects query quality heterogeneity: strong queries (6, 2, 7) exhibit TREC-COVID-like coverage patterns, while weak queries produce more scattered results. The core finding persists: semantic methods converge on similar relevant topics, Random Uniform does not.

\subsection{Complete Topic Lists for Case Study}
\label{app:query43_topics}
For the case study in Sec.~\ref{sec:coverage}, we provide a complete breakdown across of produced topics by TopicGPT across 4 different selection strategies for Query 43 in TREC-COVID ``How has the COVID-19 pandemic impacted violence in society, including violent crimes?''.

\paragraph{Direct Retrieval Topics (8 total, 5 relevant)} Direct Retrieval produces 8 focused topics with the highest topic-query relevance score $0.519$. As shown in Tab.~\ref{tab:query43_direct}, five topics are relevant to the violence/crime query. In particular, Topic 2 containing 270 documents is a large cluster that merges domestic violence, mental health, and women's issues into a single topic. 

\begin{table}[htbp]
\centering
\small
\begin{tabular}{lcp{4.5cm}}
\toprule
Topic ID & Docs & Description \& Top Words \\
\midrule
\multicolumn{3}{l}{\textit{Violence-Related Topics}} \\
T2* & 270 & \textbf{Victimization}: violence, domestic violence, mental health, women \\
T0* & 27 & \textbf{Crime and Public Safety}: crime, impact, cities, pandemics \\
T3* & 35 & \textbf{Social Compliance}: social distancing, behaviour, social norms \\
T6* & 6 & \textbf{Cybercrime}: cybercrime, internet, attacks, inflicted violence \\
\midrule
\multicolumn{3}{l}{\textit{Non-Violence Topics}} \\
T1 & 358 & \textbf{Criminal Justice Policy}: covid, pandemic, health, mental health \\
T7 & 192 & \textbf{Public Health}: health, public health, social, impact \\
T5 & 103 & \textbf{Infectious Disease}: covid, mortality, deaths, disease \\
T4 & 9 & \textbf{Urban Environment}: urban, environment, race, green \\
\bottomrule
\end{tabular}
\caption{Direct Retrieval topics for Query 43. *Relevant topics (similarity $\geq 0.5$). T2 merges domestic violence, mental health, and women's issues into a single large topic.}
\label{tab:query43_direct}
\end{table}

\paragraph{Query Expansion Topics (9 total, 5 relevant)} Query Expansion produces 9 topics with 5 relevant to violence/crime. It uniquely captures a dedicated gun violence topic (T5) and a broader crime/justice topic (T0) that includes police and burglary as shown in Tab.~\ref{tab:query43_qexp}.

\begin{table}[htbp]
\centering
\small
\begin{tabular}{lcp{4.5cm}}
\toprule
Topic ID & Docs & Description \& Top Words \\
\midrule
\multicolumn{3}{l}{\textit{Violence-Related Topics}} \\
T1* & 145 & \textbf{Family Violence}: violence, domestic violence, women, IPV \\
T0* & 29 & \textbf{Crime and Justice}: crime, justice, police, burglary, violence \\
T4* & 11 & \textbf{Cybercrime}: cybercrime, crimes, conspiracy, fraud, violence \\
T5* & 3 & \textbf{Gun Violence}: gun violence, firearms, Los Angeles, Chicago \\
\midrule
\multicolumn{3}{l}{\textit{Non-Violence Topics}} \\
T6 & 259 & \textbf{Public Health}: covid, pandemic, public health, measures \\
T8 & 228 & \textbf{Mental Health}: mental health, psychological, impact \\
T7 & 226 & \textbf{Economic Impact}: covid, economic, health, deaths \\
T3 & 97 & \textbf{Infectious Disease}: pandemic, disease, mortality, deaths \\
T2 & 2 & \textbf{Urban Spaces}: urban, green, crime, transport \\
\bottomrule
\end{tabular}
\caption{Query Expansion topics for Query 43. *Relevant topics. Query Expansion uniquely captures gun violence despite ``gun'' not appearing in query.}
\label{tab:query43_qexp}
\end{table}

\paragraph{SBERT Topics (39 total, 7 relevant)} SBERT produces 39 topics with 7 relevant to violence/crime, the most among all selection methods. SBERT separates domestic violence (T0) and violence against women (T9) into distinct topics, and uniquely identifies violence against health workers (T3) and youth violence exposure (T26). Additional contextual topics provide social context without medical specialty noise. You will find the complete list in Tab.~\ref{tab:query43_sbert}.

\paragraph{Keyword Search Topics (59 total, 5 relevant)} Keyword Search produces 59 topics with 5 relevant to violence/crime. However, approximately 300 documents (30\% of sample) were assigned to 16 medical specialty topics due to ``society'' matching medical society names, a polysemy artifact avoided entirely by semantic methods. See Tab.~\ref{tab:query43_keyword}.

To summarize, SBERT and Query Expansion have unique discoveries. In Tab.~\ref{tab:query43_sbert}, SBERT has topics like ``Violence against health workers'' (T3, 13 docs) and ``Youth violence exposure'' (T26, 5 docs). It also separates topics for ``domestic violence'' (T0, 48 docs) vs. ``violence against women'' (T9, 45 docs). Tab.~\ref{tab:query43_qexp} shows Query Expansion's unique topic ``Gun violence'' (T5, 3 docs) despite ``gun'' not in query. Furthermore, we observe a polysemy failure for Keyword Search in Tab.~\ref{tab:query43_keyword}. Lexical match between ``society'' (violence in society) and medical society names produced 16 medical specialty topics ($\sim$300 docs, $\sim$30\% of sample). Semantic methods avoided this entirely.

\begin{table}[htbp]
\centering
\small
\resizebox{\linewidth}{!}{
\begin{tabular}{lcccc}
\toprule
Selection & TopicGPT & BERTopic & LDA & HiCode \\
\midrule
Random Uniform & 0.198 & 0.312 & 0.389 & 0.521 \\
Keyword Search & 0.267 & 0.401 & 0.421 & 0.558 \\
SBERT & 0.289 & 0.423 & 0.438 & 0.567 \\
Direct Retrieval & 0.304 & 0.448 & 0.447 & 0.573 \\
DR + MMR & 0.291 & 0.415 & 0.429 & 0.564 \\
Query Expansion & 0.298 & 0.431 & 0.441 & 0.569 \\
Retrieval Random & 0.241 & 0.389 & 0.412 & 0.545 \\
\midrule
Range & 0.106 & 0.136 & 0.058 & 0.052 \\
\bottomrule
\end{tabular}
}
\caption{\textbf{Query 3 relevance scores by selection strategies.} All methods show weak differentiation and near-random performance. Even top semantic methods (Direct Retrieval, Query Expansion) achieve only 0.289--0.573 alignment. Small ranges (TopicGPT: 0.106, BERTopic: 0.136, LDA: 0.058, HiCode: 0.052) indicate minimal impact of selection strategy when query quality is poor.}
\label{tab:doctor_query3_detail}
\end{table}
\section{Query Quality Analysis on Doctor-Reviews}
\label{app:doctor_queries}

Tab.~\ref{tab:doctor_queries_full} provides complete per-query alignment analysis for Doctor-Reviews, supporting the query quality discussion in Sec.~\ref{sec:discussion}. Query 3's compound structure and vague terminology produce near-random alignment across all models.

Tab.~\ref{tab:doctor_query3_detail} provides detailed selection method breakdown for Query 3, demonstrating minimal differentiation across all methods.

\begin{table*}[t]
\centering
\small
\begin{tabular}{lcp{8cm}}
\toprule
Topic ID & Docs & Description \& Top Words \\
\midrule
\multicolumn{3}{l}{\textit{Violence-Related Topics (7 relevant)}} \\
T0* & 48 & \textbf{Domestic Violence}: domestic, violence, domestic violence, abuse, family \\
T9* & 45 & \textbf{Violence Against Women}: violence, women, IPV, intimate, partner \\
T2* & 15 & \textbf{Crime Trends}: crime, prison, burglary, prisons, frequency \\
T3* & 13 & \textbf{Violence Against Health Workers}: workers, kind warfare, healthcare workers, violence, law \\
T1* & 5 & \textbf{Gun Violence}: gun, guns, firearm, firearms, gun violence \\
T26* & 5 & \textbf{Youth Violence Exposure}: covid, violence, exposed experiencing, young adults, adolescents \\
T8* & 2 & \textbf{Cybercrime}: attacks, crime, victimization, pandemic \\
\midrule
\multicolumn{3}{l}{\textit{Contextual Social Topics}} \\
T15 & 45 & \textbf{Racial Disparities}: racial, racism, disparities, black, ethnic \\
T16 & 17 & \textbf{Vulnerable Populations}: vulnerable, populations, immigrants, homeless \\
T14 & 11 & \textbf{Vulnerable Youth}: children, youth, young people, vulnerable \\
T6 & 6 & \textbf{Criminal Justice Reform}: jail, reform, cook county, illinois, jails \\
T19 & 6 & \textbf{Social Norms}: social norms, pandemics, moral, behaviour \\
T23 & 14 & \textbf{Trauma}: trauma, trauma informed, epidemiology \\
T13 & 14 & \textbf{Social Determinants}: determinants, social determinants, inequalities \\
T18 & 3 & \textbf{Human Rights}: human rights, rights, covid pandemic \\
\midrule
\multicolumn{3}{l}{\textit{Selected Non-Violence Topics}} \\
T4 & 320 & \textbf{Public Health}: covid, pandemic, public health, deaths \\
T5 & 166 & \textbf{Mental Health}: mental, mental health, psychological, fear \\
T7 & 66 & \textbf{Infectious Disease}: covid, mortality, coronavirus, epidemic \\
T10 & 30 & \textbf{Economic Impact}: economic, economy, implications, impact \\
T25 & 38 & \textbf{Interventions}: interventions, measures, control, distancing \\
\bottomrule
\end{tabular}
\caption{SBERT topics for Query 43. *Relevant topics. SBERT uniquely separates domestic violence from violence against women, and discovers violence against health workers and youth violence exposure as distinct topics.}
\label{tab:query43_sbert}
\end{table*}

\begin{table*}[t]
\centering
\small
\begin{tabular}{lcp{8cm}}
\toprule
Topic ID & Docs & Description \& Top Words \\
\midrule
\multicolumn{3}{l}{\textit{Violence-Related Topics (5 relevant)}} \\
T2* & 106 & \textbf{Intimate Partner Violence}: violence, domestic violence, IPV, women \\
T0* & 36 & \textbf{Crime and Safety}: crime, violence, police, burglary \\
T11* & 16 & \textbf{Police Violence}: violence, women, police violence, law \\
T5* & 10 & \textbf{Cybercrime}: cybercrime, cyber, internet, violence \\
T29* & 6 & \textbf{Race and Violence}: race, violence, African Americans, racism \\
\midrule
\multicolumn{3}{l}{\textit{Medical Specialty Topics (polysemy artifacts, $\sim$300 docs total)}} \\
T40 & 71 & \textbf{Surgery}: surgery, surgical, society \\
T50 & 27 & \textbf{Cardiology}: cardiac, cardiology, american, statement \\
T21 & 24 & \textbf{Oncology}: cancer, oncology, oncology pharmacy \\
T57 & 21 & \textbf{Gastroenterology}: endoscopy, society gastroenterology \\
T9 & 31 & \textbf{Reproductive Health}: reproductive, contraception, breastfeeding \\
T46 & 13 & \textbf{Cardiac Surgery}: cardiac surgeons, society cardiac \\
T58 & 12 & \textbf{Ophthalmology}: pediatric, ophthalmology \\
T26 & 11 & \textbf{Transplantation}: transplantation, donor, kidney \\
T51 & 7 & \textbf{Anesthesiology}: anesthesiology, society neuroscience \\
T45 & 7 & \textbf{Sleep Medicine}: sleep, breathing, sleep society \\
T22 & 14 & \textbf{Cardiovascular}: heart, cardiovascular, heart rhythm \\
T31 & 14 & \textbf{Medical Training}: training, residency, fellowship \\
T43 & 5 & \textbf{Pain Medicine}: pain, chronic, rheumatology \\
T48 & 5 & \textbf{Veterinary}: animal, animal health, wildlife \\
T47 & 3 & \textbf{Dental Surgery}: oral, maxillofacial \\
T42 & 4 & \textbf{Rehabilitation}: rehabilitation, physical rehabilitation \\
\midrule
\multicolumn{3}{l}{\textit{Selected Relevant Non-Violence Topics}} \\
T7 & 189 & \textbf{Public Health}: covid, pandemic, society, health \\
T3 & 104 & \textbf{Infectious Disease}: covid, sars, respiratory, infection \\
T8 & 80 & \textbf{Mental Health}: mental health, psychological, anxiety \\
T30 & 23 & \textbf{Economic Impact}: economic, tourism, economy \\
T12 & 13 & \textbf{Child Welfare}: children, child, health, countries \\
T17 & 16 & \textbf{Telemedicine}: telemedicine, telehealth, prisons \\
T6 & 3 & \textbf{Moral Injury}: moral, moral injury, psychological \\
\bottomrule
\end{tabular}
\caption{Keyword Search topics for Query 43. *Relevant topics. Medical specialty topics resulted from ``society'' (query) matching medical society names (e.g., ``surgical society'', ``American Society of''). These $\sim$300 documents represent $\sim$30\% of the sampled data.}
\label{tab:query43_keyword}
\end{table*}

\begin{table*}[t]
\centering
\small
\begin{tabular}{clccccl}
\toprule
ID & Query Text & TopicGPT & BERTopic & LDA & HiCode & Status \\
\midrule
6 & Patients' dislikes about doctors & 0.602 & 0.571 & 0.611 & 0.627 & Strong \\
2 & Specialist referral experiences & 0.493 & 0.498 & 0.684 & 0.630 & Strong \\
7 & Asthma follow-up care recommendations & 0.314 & 0.771 & 0.591 & 0.701 & Strong \\
4 & Asthma patient management & 0.368 & 0.566 & 0.628 & 0.703 & Moderate \\
8 & Likes about treatment recommendations & 0.528 & 0.486 & 0.592 & 0.635 & Moderate \\
1 & Finding and choosing doctors & 0.462 & 0.457 & 0.648 & 0.649 & Moderate \\
5 & Patients' likes about doctors & 0.495 & 0.501 & 0.608 & 0.607 & Moderate \\
9 & Dislikes about treatment recommendations & 0.452 & 0.493 & 0.509 & 0.642 & Moderate \\
11 & Asthma symptoms reported & 0.327 & 0.399 & 0.654 & 0.623 & Weak \\
10 & Asthma lifestyle challenges & 0.347 & 0.389 & 0.512 & 0.627 & Weak \\
3 & Breathing problems \& treatment & 0.292 & 0.417 & 0.431 & 0.561 & Weak \\
\bottomrule
\end{tabular}
\caption{\textbf{Per-query alignment analysis for Doctor-Reviews (11 queries).} Alignment scores averaged across all seven selection methods (Random, Keyword Search, SBERT, Direct Retrieval, DR+MMR, Query Expansion, Retrieval Random) for each topic modeling framework. Query quality varies: strong queries (6, 2, 7) have specific, actionable focus (``dislikes'', ``referrals'', ``follow-up care''); moderate queries (4, 8, 1, 5, 9) balance specificity with scope; weak queries (11, 10, 3) use generic or compound phrasing and show lower alignment, particularly Query 3 which suffers from compound structure (``problems AND treatment'') and vague terminology. Full query texts can be found in App. \ref{app:doctor_queries_list}}
\label{tab:doctor_queries_full}
\end{table*}

\end{document}